\documentclass[usenatbib]{mn2e}
\usepackage{multirow}
\usepackage{rotating}
\usepackage{colortbl}
\usepackage{color}
\usepackage[fleqn]{amsmath}
\usepackage{amssymb}
\usepackage{amsfonts}
\usepackage{verbatim}
\usepackage{scalefnt}
\usepackage{lscape}
\usepackage{booktabs}

\usepackage[bookmarks=False,bookmarksnumbered,colorlinks=true, citecolor=blue, linkcolor=cyan]{hyperref}

\newcommand{\apjl}{ApJ}
\newcommand{\apjs}{ApJS}
\newcommand{\ao}{Appl.~Opt.}
\newcommand{\aap}{A\&A}

\newcommand{\beq}{\begin{equation}}
\newcommand{\eeq}{\end{equation}}

\title[Faint high-redshift quasars and JWST]
{Co-evolution of massive black holes and their host galaxies at high redshift: discrepancies from six cosmological simulations and the key role of JWST}
\author[Habouzit et al.]{M\'elanie Habouzit$^{1,2}$\thanks{E-mail: habouzit@mpia.de}, Masafusa Onoue$^2$, Eduardo Ba\~nados$^2$, 
Marcel Neeleman$^2$,
\newauthor Daniel Angl\'es-Alc\'azar$^{3,4}$, Fabian Walter$^2$, Annalisa Pillepich$^2$, Romeel Dav\'e$^{5}$,
\newauthor Knud Jahnke$^2$, and Yohan Dubois$^{6}$\\
\\
$^1$ Zentrum für Astronomie der Universit\"at Heidelberg,
 ITA, Albert-Ueberle-Str. 2, D-69120 Heidelberg, Germany\\
$^2$ Max-Planck-Institut f\"ur Astronomie, K\"onigstuhl 17, D-69117 Heidelberg, Germany\\
 $^{3}$ Department of Physics, University of Connecticut, 196 Auditorium Road, U-3046, Storrs, CT 06269-3046, USA\\
 $^4$ Center for Computational Astrophysics, Flatiron Institute, New York, NY 10010, USA\\
 $^{5}$ Institute for Astronomy, Royal Observatory, University of Edinburgh, Edinburgh EH9 3HJ, UK\\ 
 $^{6}$ Institut d'Astrophysique de Paris, Sorbonne Universit\'es, CNRS, UMR 7095, 98 bis bd Arago, 75014 Paris, France
}

\date{2021}                  

\begin{document}
\maketitle

\begin{abstract}
The James Webb Space Telescope will have the power to characterize high-redshift quasars at $z\geqslant 6$ with an unprecedented depth and spatial resolution. While the brightest quasars at such redshift (i.e., with bolometric luminosity $L_{\rm bol}\geqslant 10^{46}\, \rm erg/s$) provide us with key information on the most extreme objects in the Universe, measuring the black hole (BH) mass and Eddington ratios of fainter quasars with $L_{\rm bol}= 10^{45}-10^{46}\, \rm erg/s$ 
opens a path to understand the build-up of more {\it normal} BHs at $z\geqslant 6$. 
In this paper, we show that the Illustris, TNG100, TNG300, Horizon-AGN, EAGLE, and SIMBA large-scale cosmological simulations do not agree on whether BHs at $z\geqslant 4$ are overmassive or undermassive at fixed galaxy stellar mass with respect to the $M_{\rm BH}-M_{\star}$ scaling relation at $z=0$ (BH mass offsets). Our conclusions are unchanged when using the local scaling relation produced by each simulation or empirical relations. 
We find that the BH mass offsets of the simulated {\it faint quasar} population at $z\geqslant 4$, unlike those of bright quasars, represent the BH mass offsets of the entire BH population, for all the simulations.
Thus, a population of {\it faint quasars} with $L_{\rm bol}= 10^{45}-10^{46}\, \rm erg/s$ observed by JWST can provide key constraints on the assembly of BHs at high redshift.
Moreover, this will help constraining the high-redshift regime of cosmological simulations, including BH seeding, early growth, and co-evolution with the host galaxies. 
Our results also motivate the need for simulations of larger cosmological volumes down to $z\sim 6$, with the same diversity of sub-grid physics, in order to gain statistics on the most extreme objects at high redshift.

\end{abstract}

\section{Introduction}

The empirical scaling relation between massive black hole (BH) mass and the stellar mass of their host galaxies ($M_{\rm BH}-M_{\star}$) found in the local Universe \citep[e.g.,][]{Magorrian1998,2009ApJ...698..198G} 
indicates that, on average, more massive galaxies host more massive BHs. Other scaling relations have been derived between BH mass and galaxy properties such as galaxy bulge mass, velocity dispersion, Sérsic index, and infrared luminosity \citep[e.g.,][]{MarconiHunt2003,Haring2004,Shankar2004,2015ApJ...798...54G}.
These relations could represent the evidence for the cosmic co-evolution of BHs with their host galaxies.
The $M_{\rm BH}-M_{\star}$ relation could emerge from the hierarchical build-up of galaxies:  
BHs could merge (if BH coalescence is an efficient process) and increase their mass along the scaling relation after the mergers of their host galaxies.
As the build-up of such relations is not constrained yet in the high-redshift Universe, other pathways are possible: BH growth could precede the assembly of their host galaxies, or galaxies could grow first and their BHs would only catch up later on.
Theoretically, the co-evolution between BHs and their host galaxies is predicted to be complex as the shape, normalization and scatter of the $M_{\rm BH}-M_{\star}$ relation can be impacted by several key physical processes related to BH and galaxy evolution, such as accretion onto the BHs, and feedback processes from supernovae (SN) and AGN \citep[][from the perspective of cosmological simulations]{2021MNRAS.503.1940H}.  
The massive end of the scaling relation can be reproduced with a feedback-regulated model of BH growth \citep[e.g.,][]{DiMatteo2005,Dimatteo2008,2012MNRAS.420.2662D}, but can also be obtained without invoking AGN feedback \citep{2013ApJ...770....5A}.
The shape of the relation, and the low-mass end of the relation, is influenced by the ability of SN feedback to regulate the growth of BHs \citep[e.g.,][]{2015MNRAS.452.1502D,
2017MNRAS.468.3935H,2017MNRAS.472L.109A,2018MNRAS.481.3118M}.
BH seeding and dynamics also likely play a role: BH seeds with small initial mass have a hard time sinking to the center of their host galaxies, being off of the gas reservoir they would not be in the ideal position to accrete gas efficiently \citep[e.g.,][]{2017MNRAS.472L.109A,2019MNRAS.486..101P,2020arXiv200712185C,2021arXiv210102727M}.
Some models based only on BH mergers (i.e., without considering gas accretion or for which only a small effect was found) were also able to reproduce local scaling relations \citep{Peng2007,2010MNRAS.407.1016H,2011ApJ...734...92J}.
The establishment of the scaling relations is still unclear, and poorly constrained with observations as processes of gas accretion, BH mergers, and feedback remain hard to quantify. \\
\indent Whether the $M_{\rm BH}-M_{\star}$ relation evolves toward high redshift is a key question, but selection biases in observations make the investigation difficult. 
Current constraints at $z\leqslant 2$ are mostly consistent with mild to no evolution or showing slightly higher $M_{\rm BH}/M_{\star}$ ratios at higher redshift \citep[e.g.,][]{2003ApJ...583..124S,
2009ApJ...706L.215J,
2019arXiv191202824S}. 
Measuring stellar mass requires sufficient sensitivity and spatial resolution to remove the extremely bright nuclear emission. 
Such observations are challenging beyond $z\sim 1.5$ where the redshifted $4000\ \rm \AA$ break falls beyond $\rm 1\ \mu m$ \citep[e.g.,][]{2009ApJ...706L.215J,2020ApJ...888...37D}, even with current ground-based facilities and the optical-sensitive Hubble Space Telescope (HST).

\begin{figure*}
    \centering
    \includegraphics[scale=0.6]{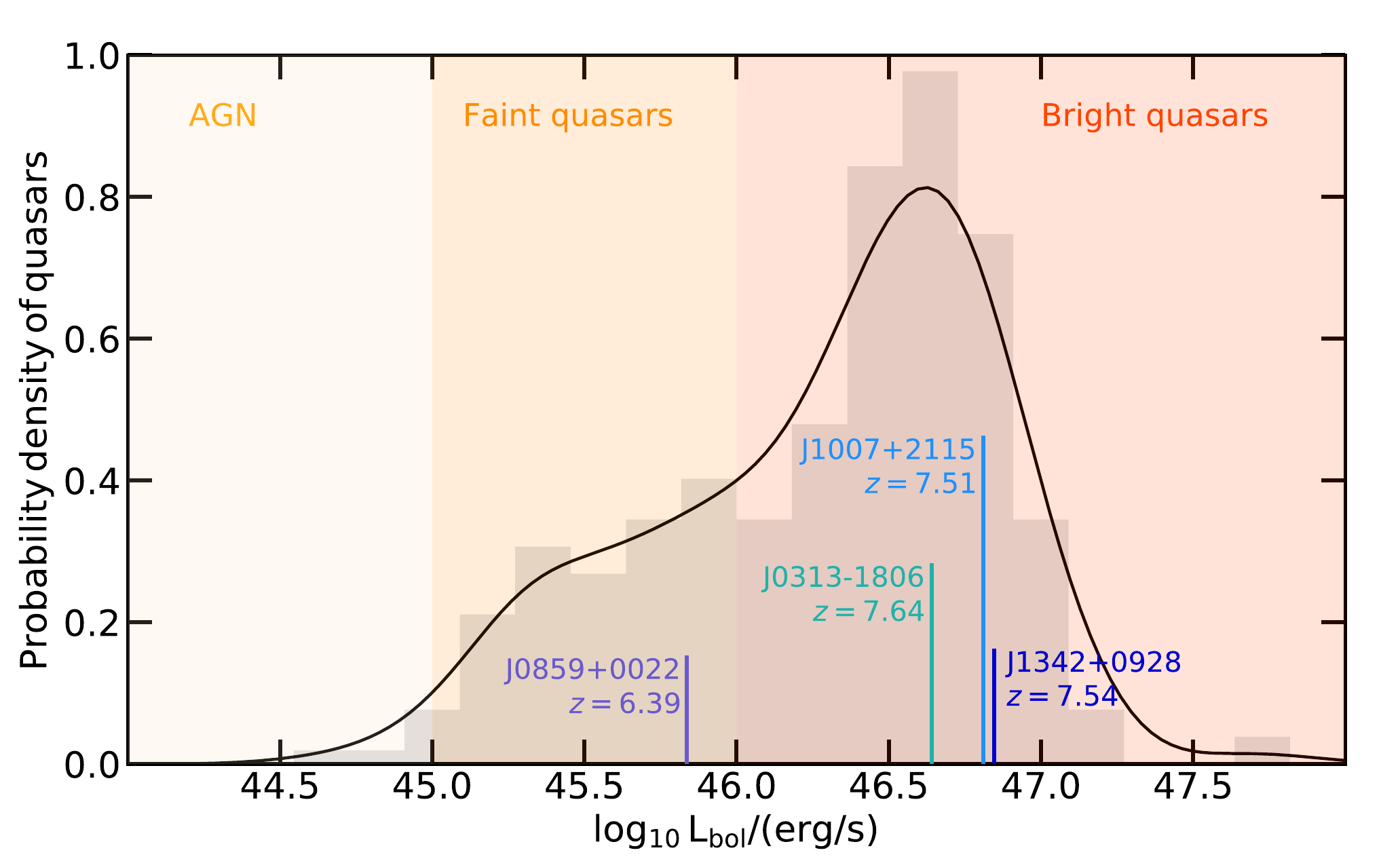}
    \caption{The probability density of the 287 quasars at $z\geqslant 5.8$ that have been published up to March 2021. 
    For reference, we indicate the bolometric luminosity of the three highest-redshift quasars detected beyond $z=7.5$ \citep{2018ApJ...856L..25B,2020ApJ...897L..14Y,2021arXiv210103179W} and the least luminous $z>5.8$ quasar with an estimate of its black hole mass to date \citep{2019ApJ...880...77O}. In the following we define three categories of quasars: the bright quasars with $L_{\rm bol}\geqslant 10^{46}\, \rm erg/s$, the faint quasars with $L_{\rm bol}\geqslant 10^{45}-10^{46}\, \rm erg/s$, and fainter objects with $L_{\rm bol}\leqslant 10^{45}\, \rm erg/s$ that we refer to as AGN.}
    \label{fig:300quasars}
\end{figure*}

In this paper, we analyze the most massive and active BHs produced by the Illustris \citep{2014MNRAS.445..175G,2014MNRAS.444.1518V,2015MNRAS.452..575S}, TNG100, TNG300 \citep{2017arXiv170302970P,2017arXiv170703397S,2018MNRAS.475..624N,2017arXiv170703401N,2018MNRAS.480.5113M,2015A&C....13...12N}, Horizon-AGN \citep{2014MNRAS.444.1453D,2016MNRAS.463.3948D,2016MNRAS.460.2979V}, EAGLE \citep{2015MNRAS.446..521S,2015MNRAS.450.1937C,2016A&C....15...72M}, and SIMBA \citep{2019MNRAS.486.2827D} simulations. 
Those can be referred as quasars. Quasars are the most luminous class of AGN, and a phase of rapid evolution through gas accretion. They are powered by BHs with $M_{\rm BH}\sim 10^{8}-10^{10}\, \rm M_{\odot}$, rivaling the most massive BHs in the local Universe \citep[e.g.,][]{Mortlock2011,2018ApJ...856L..25B,2021arXiv210103179W,2020ApJ...897L..14Y}.
To date, 287 quasars at $z\geqslant 5.8$ have been published\footnote{We compiled a list of all published $z\geqslant 5.8$ quasars up to March 2021: \citet{2001AJ....122.2833F,2004AJ....128..515F,
2006AJ....131.1203F,2006AJ....132..823C,
2006MNRAS.371..769G,2006ApJ...652..157M,
2009AJ....138..305J,2009A&A...505...97M,
2009AJ....137.3541W,2010AJ....139..906W,
Mortlock2011,2011ApJ...736...57Z,
2011ApJ...739...56D,2015ApJ...801L..11V, 2015AJ....149..188J,2015ApJ...813L..35K,
2015MNRAS.451L..16C,2015ApJ...798...28K, 2016ApJS..227...11B,2016ApJ...833..222J,
2016ApJ...828...26M,2017MNRAS.468.4702R,
2017ApJ...834...83M,2017MNRAS.466.4568T,2018Natur.553..473B,
2018ApJ...854...97D,2018ApJ...861L..14B,
2018ApJ...869L...9W,2018MNRAS.478.1649C,2019ApJ...872L...2M,
2019MNRAS.487.1874R,2020ApJ...897L..14Y,2020ApJ...903...34A,
2021ApJ...909...80B,2021ApJ...907L...1W}.}, 
among which $\sim$$50$ have BH mass measurements from modeling their broad emission lines \citep[e.g.,][]{2019ApJ...873...35S,2020ApJ...905...51S}. We show the distribution of their bolometric luminosities in Fig.~\ref{fig:300quasars}.
For the purpose of our analysis we divide the quasars in three categories: {\it bright quasars} with $L_{\rm bol}\geqslant 10^{46}\rm \, erg/s$, {\it faint quasars} with $L_{\rm bol}= 10^{45}-10^{46}\rm \, erg/s$\footnote{Our definition of {\it faint quasars} relies on the faintest quasars that have been observed so far.}, 
and fainter objects with $L_{\rm bol}\leqslant 10^{45}\rm \, erg/s$ are referred as AGN in the following.
{\it Faint quasars} are particularly promising laboratories to confront theory with observations: they are produced in sufficient numbers in cosmological simulations, but also have been detected through deep quasar surveys, and will continue to be detected in the near future with new facilities.

Now is the perfect time to investigate the build-up of the $M_{\rm BH}-M_{\star}$ relation
at $z\geqslant 5$ for two reasons: the James Webb Space Telescope (JWST) should be able to constrain the stellar component of high-redshift quasar host galaxies 
(i.e., measuring the stellar mass and not just the galaxy dynamical mass, as explained below), and
to characterize previously identified {\it faint} quasars, which is not possible with current facilities.
As shown in this paper, {\it faint} quasars with $L_{\rm bol}=10^{45}-10^{46}\, \rm erg/s$ could have properties more representative of the entire BH population than bright quasars with $L_{\rm bol}\geqslant 10^{46}\rm \, erg/s$.

Measuring the stellar mass of galaxies requires observations at rest-frame UV and optical, which correspond to the near-infrared for galaxies at $z\geqslant 6$. This has proven to be extremely challenging and thus far impossible for $z\geqslant 6$ quasar host galaxies 
\citep[e.g.][]{2012ApJ...756..150D,2020ApJ...900...21M}.   
Currently, our knowledge of $z\geqslant 6$ quasar hosts come only from rest-frame far-infrared emission \citep[e.g.,][]{2004ApJ...615L..17W,
2020ApJ...904..130V,2021arXiv210509958P}, which traces the gas and cold dust component. The latter can be investigated with  
facilities such as 
ALMA, and can provide measurement for the dynamical mass of the galaxies \citep[e.g.,][]{2020A&A...637A..84P,2021arXiv210205679N}. 
The JWST will enable the characterization of the stellar component of high-redshift quasar hosts for the first time \citep[e.g.,][]{2020MNRAS.499.3819M,2021arXiv210101219M}.
Still, measuring the stellar mass will be challenging even with JWST. For example, separating the emission from the quasar and the stellar component will be difficult if the latter is compact, as it has been shown in some cases with ALMA dust and [C{\sc ii}] 158$\mu$m measurements \citep{2017ApJ...837..146V,2019ApJ...882...10N,2020ApJ...904..130V}.

We also expect the high NIR sensitivity of JWST to allow us to measure BH mass for the {\it faint quasars}, which are too faint to perform NIR spectroscopy with ground-based 8m-class telescopes in a reasonable amount of time (see e.g., \citealt{2010AJ....140..546W,2018ApJ...855..138K,2019ApJ...880...77O}).  
NIR (rest-frame UV/optical) spectroscopy is necessary to detect BH broad-line region emission lines such as MgII, CIV, and Balmer lines, and their underlying quasar continua.
Moreover, the BH mass measurements at $z\gtrsim 4$ have relied on Mg{\sc ii} $\lambda 2798$, although the empirically calibrated Mg{\sc ii}-based MBHs have a half dex of systematic uncertainties \citep{2013BASI...41...61S}.
The JWST provides an opportunity to do rest-frame optical mass measurements with H$\beta$, which is directly calibrated against reverberation mapping measurements \citep{2006ApJ...641..689V}.

The quest for fainter quasars has already started. At the time of writing, 83 faint quasars with bolometric luminosity of $L_{\rm bol}=10^{45}-10^{46}\,\rm erg/s$ (i.e., down to $M_{\rm 1450}\sim -24$) have been identified in the redshift range $z=6-7$ \citep{2007AJ....134.2435W,2009AJ....137.3541W,2010AJ....139..906W,2013A&A...557A..78M,2015ApJ...813L..35K,2016ApJ...828...26M,2018ApJS..237....5M,2019ApJ...883..183M,2019ApJ...872L...2M}.
While surveys such as SDSS and PS1 are limited to the detection of bright quasars ($L_{\rm bol}\geqslant 10^{46}\,\rm erg/s$), the quasars with $L_{\rm bol}=10^{45}-10^{46}\,\rm erg/s$ are identified in moderately deep surveys, such as CFHQS (Canada-France High-redshift Quasar Survey) and SHELLQs (Subaru High-redshift Exploration of Low-Luminosity Quasars).
A significant fraction of those quasars were identified in the Subaru HSC-SSP project with $5\sigma$ magnitude limits of $z_{\rm AB}<24.5$ and $y_{\rm AB}<24.0$ \citep{2018PASJ...70S...8A}.
These new objects allowed the characterization of the faint end of the quasar luminosity function at $z=5$ \citep[e.g.,][]{2018AJ....155..131M,2020ApJ...904...89N} and $z=6$ \citep{2018ApJ...869..150M}, enabling estimates of the quasar luminosity function down to $M_{1450}<-22.3$.  
\citet{2019ApJ...880...77O} showed that faint $z\sim 6$ quasars were powered by BHs of with a wide range of masses ($M_{\rm BH}\sim 10^{7.5}-10^{9}\, \rm M_{\odot}$) and Eddington ratios ($f_{\rm Edd}=0.1-1$).

In this paper, we quantify the evolution of the $M_{\rm BH}-M_{\star}$ relation from $z=6$ to the local Universe with cosmological simulations, and investigate whether BHs are overmassive at high redshift.
To do so, we derive BH mass offsets, defined as $\Delta \log_{10}\,M_{\rm BH} (M_{\star},z)=\log_{10}\,M_{\rm BH} (M_{\star},z)-\log_{10}\,M_{\rm BH} (M_{\star},z=0)$.
Observational constraints on the BH mass offsets from individual $z\geqslant 6$ quasars exist in the literature, although with galaxy dynamical mass and not the galaxy stellar mass \citep[e.g.,][and references therein]{2019PASJ...71..111I}.
These works find that while the most luminous of these quasars with $M_{\rm 1450}\leqslant-25$ (i.e., $L_{\rm bol}\geqslant 10^{45.5}\, \rm erg/s$) tend to be overmassive compared to for example the $M_{\rm BH}-M_{\star}$ empirical scaling relation of \citet{2013ARA&A..51..511K} \citep[see][for a possible formation pathway of these BHs]{2021arXiv211010693I}, the faint end of the bright quasar sample with $M_{\rm 1450}\geqslant-25$ (which are on average powered by less massive BHs) tend to be consistent with the scaling relation. However, estimates of the dynamical masses convey large uncertainties \citep{2019PASJ...71..111I,2019MNRAS.488.4004L,2020A&A...637A..84P,2021arXiv210205679N}: for example if they are over-estimated, the BHs could actually lie above the scaling relation.

We describe the cosmological simulations in Section 2. In Section 3, we present the $M_{\rm BH}-M_{\star}$ relations of the simulations at high redshift, the properties of the BHs powering faint quasars, and their BH mass offsets. We discuss our results in Section 4 and conclude in Section 5.

\section{Methodology: Cosmological simulations and their BH/AGN populations}
In this work we use the following six large-scale cosmological hydrodynamical simulations: Illustris, TNG100, TNG300 (larger volume and lower resolution with respect to TNG100), Horizon-AGN, EAGLE, and SIMBA. 
All these simulations have volumes of $\geqslant 100^{3}\, \rm cMpc^{3}$, dark matter mass resolutions of $\sim 5\times 10^{6}-8\times 10^{7}\, \rm M_{\odot}$, and spatial resolutions of 1-2 ckpc. 
These simulations model the evolution of the dark matter and baryonic matter contents in an expanding space-time. 
They capture the highly non-linear processes involved in the evolution of galaxies and BHs, and spanning kpc to Mpc scales.
For physical processes taking place at small sub-galactic scale, they rely on subgrid modeling, e.g., for star formation, stellar and SN feedback, BH formation, evolution and feedback. Subgrid models vary from simulation to simulation \citep[see Section~2 of][]{2021MNRAS.503.1940H}. We summarize the simulations in Table~\ref{table:table_params}.

\subsection{Modeling of BH physics}
We briefly describe the modeling of BH seeding, growth, and AGN feedback below.
BH particles are seeded either in massive halos of $\geqslant 10^{10}\, \rm M_{\odot}$ (Illustris, TNG100, TNG300, EAGLE), or in galaxies of $M_{\star}\geqslant 10^{9.5}\, \rm M_{\odot}$ (SIMBA), or based on the properties of local gas cells (Horizon-AGN). Initial BH masses are comprised in the range $M_{\rm BH}=10^{4}-10^{6}\, \rm M_{\odot}$ ($\sim 10^{4}\, \rm M_{\odot}$ for SIMBA, $\sim 10^{5}\, \rm M_{\odot}$ for Illustris, Horizon-AGN, EAGLE, and $\sim 10^{6}\, \rm M_{\odot}$ for TNG100, TNG300).
BHs can growth by BH mergers and gas accretion.
Gas accretion is often modeled with the Bondi-Hoyle-Lyttleton formalism, with different variations: TNG100 and TNG300 include a magnetic field component \citep{2017arXiv170302970P}, EAGLE includes a viscous disk component \citep{2015MNRAS.454.1038R}. Finally, SIMBA has a two mode model for gas accretion: Bondi-Hoyle-Lyttleton model for the hot gas component ($T>10^{5}\, \rm K$), and a gravitational torque limited model for the cold gas component \citep[$T<10^{5}\, \rm K$,][]{2011MNRAS.415.1027H,2015ApJ...800..127A,2017MNRAS.464.2840A}. 
In the simulations, AGN feedback is modeled with one or two modes. A single mode is employed in EAGLE, in which thermal energy is released in the surroundings of AGN \citep{2015MNRAS.446..521S}. The other simulations use a two mode feedback. The released energy can be e.g., thermal and isotropic, and/or kinetic with collimated jets, or non-collimated outflows. In practice, the effective strength of AGN feedback varies from one simulation to another.
Illustris uses an injection of thermal energy for BHs with high accretion rates ($f_{\rm edd}\geqslant 0.05$), and also the release of thermal energy in the low accretion mode ($f_{\rm edd}\leqslant 0.05$) but as hot bubbles displaced from the BH locations \citep{2015MNRAS.452..575S}.
The TNG model uses thermal energy in the high accretion mode, and injection of kinetic energy in random directions for the low accretion mode \citep{2017MNRAS.465.3291W}. The transition between modes does not take place at a fixed Eddington ratio, but depends on BH mass as $f_{\rm Edd}=\min\left(0.002\times \left(M_{\rm BH}/10^{8}\, {\rm M_{\odot}}\right)^{2},0.1\right)$, so that on average a large fraction of BHs with $M_{\rm BH}\geqslant \rm \, a \,few\, 10^{8}\,M_{\odot}$ transition to the kinetic mode of AGN feedback \citep{2018MNRAS.479.4056W,2019MNRAS.484.4413H}. In Horizon-AGN, the high accretion mode ($f_{\rm Edd}>0.01$) also releases thermal energy isotropically, and the low accretion mode releases kinetic energy through bipolar outflows \citep{2012MNRAS.420.2662D}. In SIMBA, collimated kinetic ouflows whose velocities increase with $M_{\rm BH}$ are employed for high accretion rates, and lower mass loading factor but faster outflows for low accretion rates, whose velocities increase with decreasing $f_{\rm Edd}$. X-ray feedback is also included for SIMBA \citep{2019MNRAS.486.2827D}.

\subsection{Calibration of the simulations}
The subgrid models of the simulations are calibrated with the galaxy stellar mass function (Illustris, EAGLE, TNG, SIMBA), the galaxy size as a function of the galaxy stellar mass (EAGLE, TNG), the cosmic star formation rate density (Illustris, TNG), the stellar to halo function (Illustris, TNG), gas metallicity (Illustris), and gas fraction (TNG). All these calibrations are done by comparing the simulations to observations at $z=0$.

In addition most of these simulations are qualitatively calibrated with one of the empirical $M_{\rm BH}-M_{\star}$ scaling relations found in the local Universe. Often observed relations with the stellar mass of the galaxy bulges are used, and compared to different stellar quantities in the simulations such as the total stellar mass, the total stellar mass in the half-mass radius, or the bulge mass \citep[see section 2.6 in][]{2021MNRAS.503.1940H}. Illustris was calibrated with the $M_{\rm BH}-M_{\rm bulge}$ relation of \citet{2013ARA&A..51..511K} considering the total stellar mass within the stellar half-mass radius of the simulated galaxies as a proxy for $M_{\rm bulge}$. The TNG model was calibrated via comparison of its outcome to that of Illustris, also considering the $M_{\rm BH}-M_{\star}$ relation. 
Horizon-AGN was calibrated on the scaling $M_{\rm BH}-M_{\rm bulge}$ relation of \citet{Haring2004} using the $M_{\rm bulge}$ of the simulated galaxies. SIMBA was calibrated on the same relation assuming that the total stellar mass was a proxy for the bulge mass. EAGLE was calibrated on the \citet{McConnell2013} $M_{\rm BH}-M_{\rm bulge}$ relation, taking the stellar mass of the simulated galaxies as a proxy for the bulge mass.

The time evolution of the scaling relation was not used as a calibration in these simulations, mainly because it is poorly constrained in observations and only so at $z\leqslant 2$. The scaling relation produced in simulations at high redshift is therefore a {\it prediction} and not a direct result of calibration. As a result, it can be compared to observations to understand and constrain the co-evolution between BHs and their host galaxies.

\subsection{AGN luminosity}
In this paper, we compute the luminosity of the BHs following the model of \citet{2014MNRAS.442.2304H} \citep[built on][]{Churazov2005}, i.e. explicitly distinguishing radiatively efficient and radiatively inefficient AGN. The bolometric luminosity of radiatively efficient BHs, i.e. with an Eddington ratio of $f_{\rm Edd}=\dot{M}_{\rm BH}/\dot{M}_{\rm Edd}>0.1$, is defined as 
$L_{\rm bol}=0.1 \dot{M}_{\rm BH} c^{2}$. 
BHs with smaller Eddington ratio of $f_{\rm Edd}\leqslant 0.1$ are considered radiatively inefficient and their bolometric luminosities are computed as $L_{\rm bol}=0.1 L_{\rm Edd} (10 f_{\rm Edd})^{2}$.
For simplicity, we use the same radiative efficiency $\epsilon_{\rm r}=0.1$ for all the simulations. This is the parameter that was used in Horizon-AGN, EAGLE, and SIMBA, and $\epsilon_{\rm r}=0.2$ was used for Illustris, TNG100, and TNG300.
Since in the following we are interested in the brightest objects formed in the simulations, we do not correct the simulated BH population with a possible obscuration of the BHs. The  population of $z\geqslant 6$ quasars known to date are mostly non-obscured type I quasars \citep[see discussion in][]{2019ApJ...887..171C,2019A&A...628L...6V,2020ApJ...900..189C,2021arXiv210306901V}. This is likely due to a bias against obscured quasars in the  selection process of $z>6$ quasars, as observations have identified obscured quasars up  to $z\sim 4.6$ \citep[e.g., ][]{2015ApJ...804...27A,2020A&A...642A.149V,2021arXiv210409495D}.

\subsection{Summary of the main differences of the simulated BH and AGN populations}
As described above, the modeling of BH physics varies from one simulation to another (see Table~\ref{table:table_params}). This is also the case for the subgrid modeling of galaxy formation which can impact the BH population, such as SN feedback.
Below we briefly summarise our conclusions on the BH and AGN population of these large-scale cosmological simulations from \citet{2021MNRAS.503.1940H} and Habouzit et al., sub:
\begin{itemize}
    \item Most of the simulations produce a tight $M_{\rm BH}-M_{\star}$ relation, with a smaller $M_{\rm BH}$ intrinsic scatter at fixed galaxy stellar mass than the observed $z=0$ population.
    \item The evolution of the mean $M_{\rm BH}-M_{\star}$ relations in the redshift range $0\leqslant z \leqslant 6$ is mild, smaller than 1 dex in BH mass. 
    \item There is no consensus on the normalization and shape of the mean $M_{\rm BH}-M_{\star}$ relations in the simulations. 
    \item There is no consensus on the AGN luminosity function at high redshift in the simulations. On average, all these simulations overproduce the number of AGN with $L_{\rm bol}\leqslant 10^{45}$ at $z\sim 4$ with respect to observational constraints \citep[e.g.,][]{2010MNRAS.401.2531A,2015MNRAS.453.1946G}, but are in good agreement for more luminous objects.
 EAGLE produces the faintest population of AGN, and is in good agreement with the observations mentioned above at $z\sim4$ for the faint end of the luminosity function, but falls short for the bright end and may not produce enough AGN with $L_{\rm bol}\geqslant 10^{45}\, \rm erg/s$. 
\end{itemize}

While in broad agreement with observations, the cosmological simulations do not all have the same evolution of their galaxy population (see \citealt{2017arXiv170302970P} for the Illustris and TNG simulations). The high-redshift regime ($z\geqslant 5$) is also difficult to calibrate because the observational constraints are more uncertain, and in fact none of the simulations discussed in this work use empirical relations at high redshift to constrain the underlying physical models. In our present analysis we only discuss the relative BH to galaxy evolution, but not how well the simulated galaxies meet observational constraints.
We refer the reader to the following papers for complete analyses of the galaxy properties (e.g., galaxy morphologies, sizes, stellar mass function, UV luminosity function) of the Illustris \citep{2014Natur.509..177V,2014MNRAS.445..175G,2015MNRAS.454.1886S,2016MNRAS.458.2371R}, TNG \citep{2017arXiv170703406P,2018MNRAS.474.3976G,2018MNRAS.475..624N,2017arXiv170703397S,2020MNRAS.492.5167V,2020MNRAS.495.4747S}, Horizon-AGN \citet{2014MNRAS.444.1453D,2016MNRAS.463.3948D,2017MNRAS.tmp..224K}, EAGLE \citep{2015MNRAS.446..521S,2015MNRAS.450.1937C}, and SIMBA \citep{2019MNRAS.486.2827D,2020MNRAS.494.5636W} simulations.

\begin{figure*}
    \centering
    \includegraphics[scale=0.5]{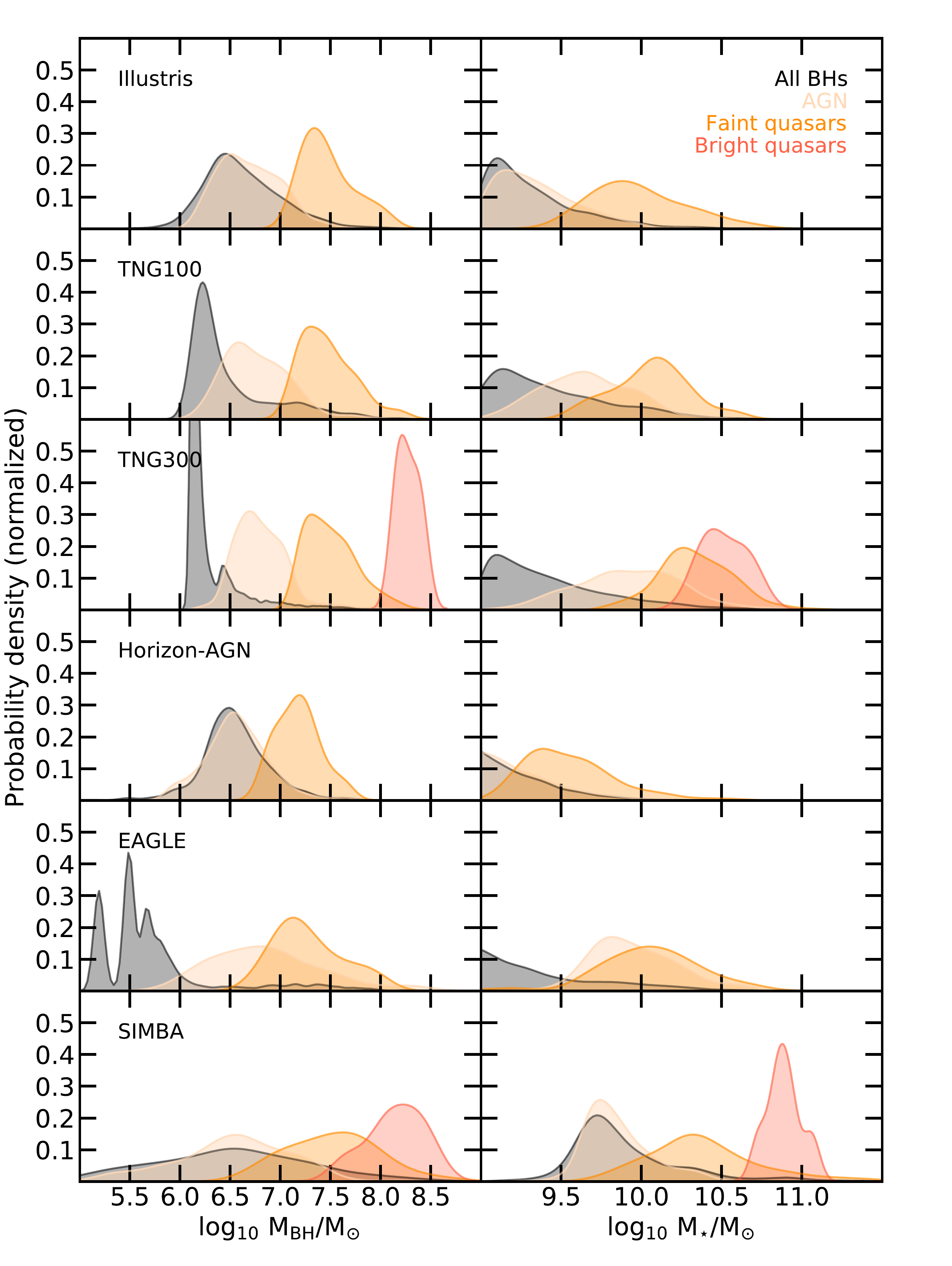}
    \caption{Normalized probability density (KDE) of BH mass (left panels) at $z=5$ and galaxy stellar mass (right panels) of all the simulated BHs (black), AGN with $\log_{10}\, L_{\rm bol}/(\rm erg/s)=44-45$ (light orange), faint quasars with $\log_{10}\, L_{\rm bol}/(\rm erg/s)=45-46$ (dark orange), and bright quasars with $\log_{10}\, L_{\rm bol}/(\rm erg/s)\geqslant 46$ (red, if enough statistics). In most of the simulations, the population of AGN is powered by BHs with mass probability densities similar to those of all the simulated population (shown in black). EAGLE has a large population of inactive BHs regulated by SN feedback, and therefore the probability density of the AGN population peaks at a higher BH mass than the entire BH population. We find a similar behavior in the TNG simulations. In general, the AGN population is representative of the entire BH population. Faint and bright quasars are powered by more massive BHs, located in more massive galaxies.
    }
    \label{fig:kde}
\end{figure*}

\begin{figure*}
    \centering
    \includegraphics[scale=1]{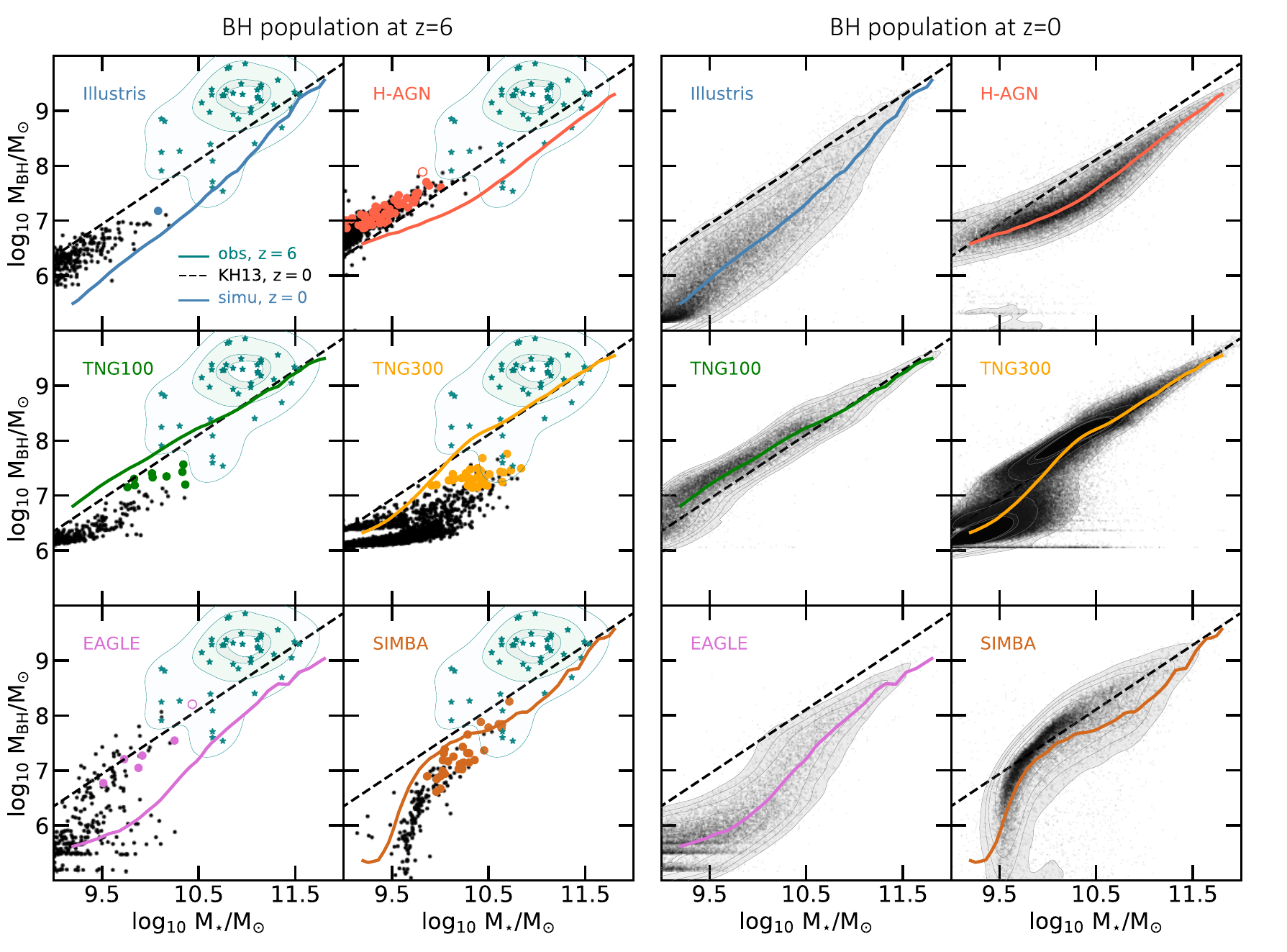}
    \caption{
    $M_{\rm BH}-M_{\star}$ plane of the simulations at $z=6$ (dot symbols, left panels) and at $z=0$ (right panels) for BHs of $M_{\rm BH}\geqslant 10^{6}\, \rm M_{\odot}$ located in galaxies of $M_{\star}\geqslant 10^{9}\, \rm M_{\odot}$. We show the $z=6$ quasars with $L_{\rm bol}\geqslant 10^{45}\, \rm erg/s$ in colored plain dots for the faint quasars and open dots for the bright quasars (only on the $z=6$ panels), while all the other BHs are shown with black dots. 
    In both sets of figures, we show the $M_{\rm BH}-M_{\star}$ median relations derived from the simulations at $z=0$ (solid colored lines). The Illustris, Horizon-AGN, and EAGLE simulations have more massive BHs at fixed stellar mass at $z=6$ than at $z=0$, and the other TNG100, TNG300, and SIMBA simulations produce the opposite behavior. The \citet{2013ARA&A..51..511K} relation in the local Universe is shown with a black dashed line in all panels, to help compare the different panels. 
    In the left panels, we also show a compilation of observed quasars at $z\gtrsim 6$ with BH mass and host dynamical mass measurements compiled in \citet{2019PASJ...71..111I} (shown as dots and probability density contours).
    }
    \label{fig:scaling}
\end{figure*}

\subsection{AGN as a proxy for the entire BH population}
When studying the average BH mass offsets, we will employ the AGN population with $L_{\rm bol}=10^{44}-10^{45}\, \rm erg/s$ (as defined in Fig.~\ref{fig:300quasars}) as a proxy for the entire BH population. 
We do so because currently only active BHs can be electromagnetically detected at high redshift.
The mass offsets derived in this paper are almost identical when considering only AGN and the entire BH population with $M_{\rm BH}\geqslant 10^{6}\, \rm M_{\odot}$.

AGN are rare objects, and thus are subdominant in terms of number density and mass contribution to the entire BH population. Still, in most of the cosmological simulations the probability densities of the AGN mass are similar to those of the BH population, as shown below.
The normalized probability density of BH mass and galaxy stellar mass of all the simulated BHs, AGN, faint quasars with $\log_{10}\, L_{\rm bol}/(\rm erg/s)=45-46$, and bright quasars with $\log_{10}\, L_{\rm bol}/(\rm erg/s)\geqslant 46$ (when enough statistics) are shown in Fig.~\ref{fig:kde} for $z=5$.
In the Illustris, Horizon-AGN, and SIMBA simulations the AGN distribution (light orange curve) resembles the full BH mass distribution (black). We note that the mass probability densities of the TNG simulations cover the same mass as the entire BH population, but are shifted towards slightly more massive BHs, and more massive host galaxies. The shift is driven by newly formed BHs, with masses of $M_{\rm BH}\sim 10^{6}\, \rm M_{\odot}$, present in abundance in the black distributions. In the absence of this peak due to BH seeds, the probability densities of the AGN and BHs are similar.
The EAGLE simulation presents the most different probability densities of the BH and AGN populations. This is due to the large number of inactive low-mass BHs in EAGLE, resulting from the efficient SN feedback and the artifical suppression of early BH growth in Bondi accretion \citep{2017MNRAS.468.3395M,2017MNRAS.465...32B}.

We use the AGN population as a tracer of the BH population in the paper and our figures below. We still confirm our results by comparing with the actual simulated BH populations, but do not display the full BH populations on the figures.

\section{Results}
\subsection{$M_{\rm BH}-M_{\star}$ scaling relations}
In Fig.~\ref{fig:scaling} we show the relation between the mass of the BHs and the total stellar mass of their host galaxies at $z=6$ (left panels) and $z=0$ (right panels), as produced by the six cosmological simulations. In each left panel, we show the full population of BHs with black dots. To help visualize the BH populations at $z=0$, we add Gaussian kernel density contours.
The AGN population with $L_{\rm bol}=10^{44}-10^{45}\, \rm erg/s$ of each simulation at $z=6$ is located in the same region as the BH population, i.e., the black dots.
At $z=6$, we show the faint and bright quasars with bolometric luminosity of $L_{\rm bol}\geqslant 10^{45}\, \rm erg/s$ with colored dots. The number density of quasars varies from one simulation to another from $n\sim 8\times 10^{-7}-1.2\times 10^{-6}\, \rm cMpc^{-3}$; we report these values in Table~\ref{tab:quasars_values}.

In some of the simulations (Illustris, TNGs), the quasars with $L_{\rm bol}\geqslant 10^{45}\, \rm erg/s$ are mainly powered by the most massive BHs at fixed galaxy stellar mass with $M_{\rm BH}\geqslant 10^{7}\, \rm M_{\odot}$. The fact that BHs less massive than $10^{7}\, \rm M_{\odot}$ do not power quasars in the TNG simulations is likely due to the efficient SN feedback model (particularly in low-mass galaxies, and at high redshift). Only TNG BHs that are massive enough ($\dot{M}_{\rm BH}\propto M_{\rm BH}^{2}$ in the Bondi formalism) and that are embedded in galaxies massive enough (e.g., $M_{\star}\geqslant 10^{9.7}\, \rm M_{\odot}$ in Fig.~\ref{fig:scaling}) to overcome SN feedback accrete sufficiently to power quasars. 
We note a larger spread of these objects at fixed $M_{\star}$ in SIMBA: some of the quasars are not among the most massive BHs at fixed $M_{\star}$, and can have masses of $M_{\rm BH}\sim 5\times 10^{6}-10^{7}\, \rm M_{\odot}$.
This is also the case for Horizon-AGN and EAGLE. The number of BHs able to power quasars, and their masses depend on the BH accretion rates (i.e., on both the accretion model, and its numerical implementation).  
While in most simulations the accretion rates scale as $M_{\rm BH}^{2}$ (Bondi), in the torque model of SIMBA the accretion rate only scales as $M_{\rm BH}^{1/6}$. Some simulations employ a boost factor, while other do not. These differences can lead a given simulation to have a general bright population of AGN (e.g., Horizon-AGN, TNGs), and another simulation to a much fainter AGN population (EAGLE and the viscous disk component of its accretion model). Some simulations use a kernel to compute the accretion rates, while others only use the BH gas cells, which can lead to more stochastic accretion rates.

The simulations do not all produce massive galaxies at the same rate, and also do not have the same simulated volume. As a result, the BHs with $L_{\rm bol}=10^{45}-10^{47}\, \rm erg/s$ are not located in the exact same galaxies. 
In Horizon-AGN, these BHs are embedded in galaxies with $M_{\star}=10^{9}-10^{10}\, \rm M_{\odot}$, the upper limit being the most massive galaxies present in the simulation at $z=6$. 
We find only one BH with $L_{\rm bol}=10^{45}-10^{47}\, \rm erg/s$ in Illustris, and it is also located in one of the most massive galaxies of the simulated volume at $z=6$ with $M_{\star}\sim 10^{10}\, \rm M_{\odot}$. The other simulations (TNGs, EAGLE, and SIMBA) host the brightest BHs in galaxies with $M_{\star}=10^{9.5}-10^{10.5}\, \rm M_{\odot}$ and even more massive galaxies for TNG300 and SIMBA.
The TNG300 simulation, which has the largest volume, had already produced galaxies of $M_{\star}\sim 10^{10.5}\, \rm M_{\odot}$ at the same time, and they host the brightest BHs.   
We add in Table~\ref{tab:quasars_values} the median, minimum and maximum values of the population of quasars with $L_{\rm bol}\geqslant 10^{45}\, \rm erg/s$ produced by the simulations.

\begin{table*}
    \centering
    \caption{Median, minimum, and maximum values of BH mass, bolometric luminosity, and host galaxy stellar mass, for BHs with $M_{\rm BH}\geqslant 10^{6}\, \rm M_{\odot}$ powering the simulated quasars with $L_{\rm bol}\geqslant 10^{45}\, \rm erg/s$ at $z=6$.}
    \begin{tabular}{l|c|c|c|c|c|c}
         & Illustris & TNG100 & TNG300 & Horizon-AGN & EAGLE & SIMBA \\
         \hline
         Number of BH(s) & 1 & 9 & 33 & 58
         & 6 & 29 \\
         Number density ($\rm cMpc^{-3}$) & $8.3 \times 10^{-7}$ & $6.6 \times 10^{-6}$ & $1.2 \times 10^{-6}$ & $2.0\times 10^{-5}$ 
         & $6.0\times 10^{-6}$ & $9.1\times 10^{-6}$\\
         \hline
         Median (min, max) & & & & & & \\
         $\log_{10}\, M_{\rm BH}/\rm M_{\odot}$  &  7.2 (-) & 7.3 (7.2,7.6) & 7.3 (7.1,7.8) & 7.2 (6.9,7.9)
         & 7.2 (6.8,8.2) & 7.2 (6.6,8.3)\\
        
         $\log_{10}\,M_{\star}/\rm M_{\odot}$ &  10.1 (-) & 10.0 (9.8,10.4) & 10.4 (9.9,10.8) & 9.3 (9.0,9.7) & 9.9 (9.5,10.4) & 10.2 (9.9,10.7)\\
         $\log_{10}\,L_{\rm bol}/\rm(erg/s)$  & 45.0 (-) & 45.2 (45.0,45.2) & 45.2 (45.0,45.6) & 45.2 (45.0,45.6) 
         & 45.2 (45.0,46.5) & 45.3 (45.0, 45.7)
    \end{tabular}

    \label{tab:quasars_values}
\end{table*}

\begin{figure*}
    \centering
    \includegraphics[scale=0.5]{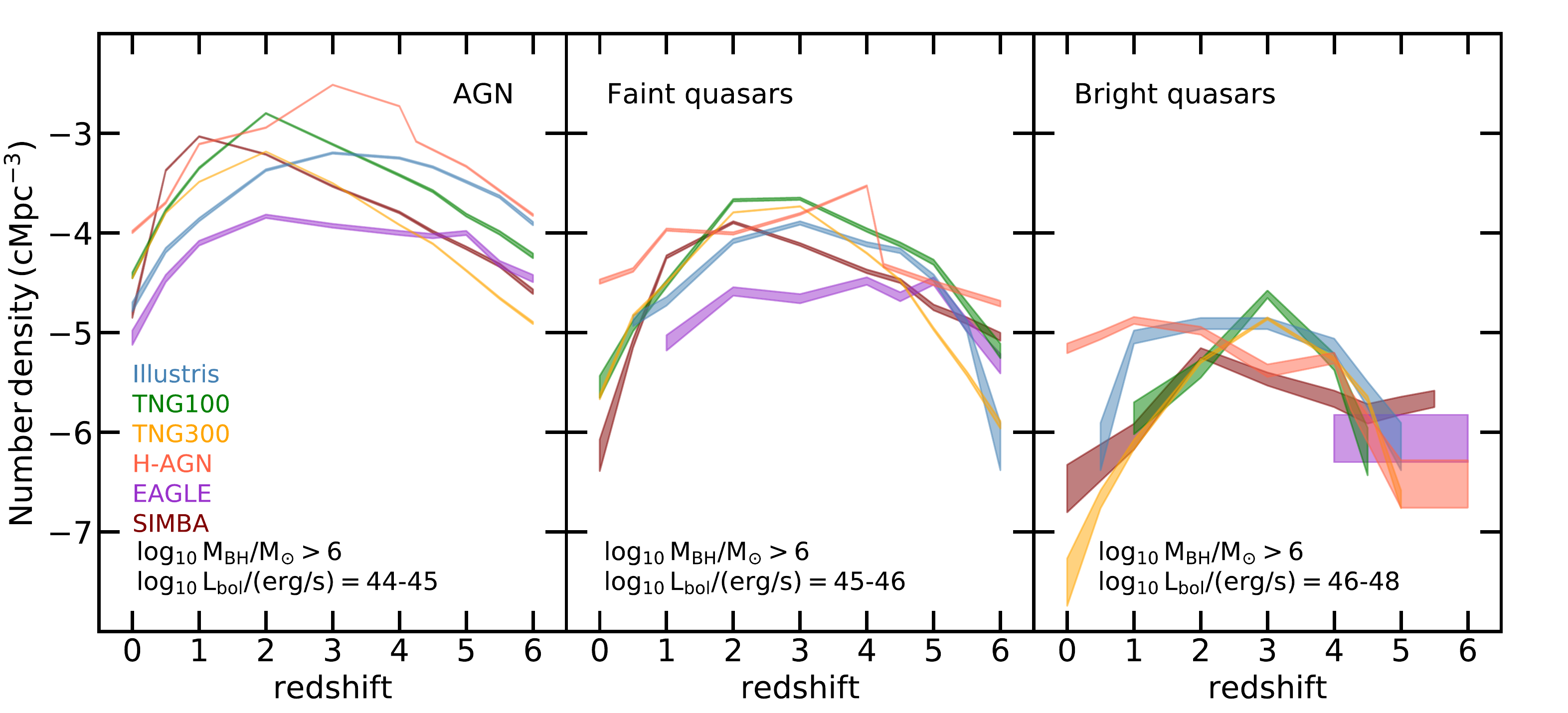}
    \caption{Redshift evolution of the comoving number density ($\rm cMpc^{-3}$, logarithmic scale) of BHs with $M_{\rm BH}\geqslant 10^{6}\, \rm M_{\odot}$ in galaxies with stellar mass of $M_{\star}\geqslant 10^{9}\,\rm M_{\odot}$. We show the AGN with bolometric lumosity $L_{\rm bol}=10^{44}-10^{45}\,\rm erg/s$ in the left panel, the {\it faint quasars} with $L_{\rm bol}=10^{45}-10^{46}\,\rm erg/s$ which could be characterized by JWST in the middle panel, and the {\it bright quasars} with $L_{\rm bol}\geqslant 10^{46}\,\rm erg/s$, similar to those of the observed $z\sim 6$ quasars, in the right panel. For the latter, we are lacking statistics given the limited simulation volumes of $100^{3}-300^{3}\, \rm cMpc^{3}$. We add Poisson error bars in all panels.     }
    \label{fig:number_density}
\end{figure*}

In Fig.~\ref{fig:scaling} we also show the mean $M_{\rm BH}-M_{\star}$ relation produced by each simulation at $z=0$ with a colored solid line. While the Illustris, Horizon-AGN, and EAGLE simulations form more massive BHs (at fixed $M_{\star}$) at $z=6$ than at $z=0$, we find that the TNG100, TNG300, and SIMBA simulations have on average less massive BHs at $z=6$. This is due to the different BH and galaxy subgrid modelings employed in all these simulations \citep{2021MNRAS.503.1940H}:

\begin{itemize}
    \item The overall normalization of the mean $M_{\rm BH}-M_{\star}$ relation decreases with decreasing redshift in Illustris, Horizon-AGN, and EAGLE. In Illustris and Horizon-AGN, this is due to a more efficient relative growth of galaxies compared to their central BHs at lower redshifts, which probably originates from a less effective SN feedback compared to other simulations. In EAGLE, SN feedback stunts the initial BH growth in low-mass galaxies, and BH rapid growth phase kicks in at fixed halo virial temperature, meaning in more massive galaxies with decreasing redshift.
    \item The overall normalization of the mean $M_{\rm BH}-M_{\star}$ relation increases with decreasing redshift in TNG100, TNG300, and SIMBA. This is due to higher BH growth at lower redshifts with respect to the BH host galaxies. In the TNG simulations this is due to a less effective SN feedback at low redshift, and in SIMBA this is mainly due to an increase of the galactic hot environment with time (due to AGN feedback), which in turn favors an additional Bondi growth channel of BHs.
\end{itemize}
 
In Fig.~\ref{fig:scaling}  we also show in light green star symbols (and the corresponding Gaussian probability density contours) the compilation of $z\geqslant 5.8$ quasars with a host dynamical mass estimate and BH mass measurement from \citet{2019PASJ...71..111I} . While faint quasars with $M_{\rm 1450}\geqslant -25$ (i.e., $L_{\rm bol}\leqslant 10^{45.5}\, \rm erg/s$) are located around the \citet{2013ARA&A..51..511K} scaling relation (when assuming $M_{\star}=M_{\rm dyn}$), the brighter quasars with $M_{\rm 1450}\leqslant -25$ tend to be overmassive compared to the same scaling relation.
For the current observations, the dynamical mass $M_{\rm dyn}$ is used, and thus we can not directly compare those to the quasars produced by the simulations for which the galaxy total stellar mass is shown. While some simulated quasars are powered by BHs with masses overlapping the observed region (i.e., TNG300, EAGLE, SIMBA), this is not the case for most of the simulated quasar populations.

\begin{figure*}
    \centering
    \includegraphics[scale=0.7]{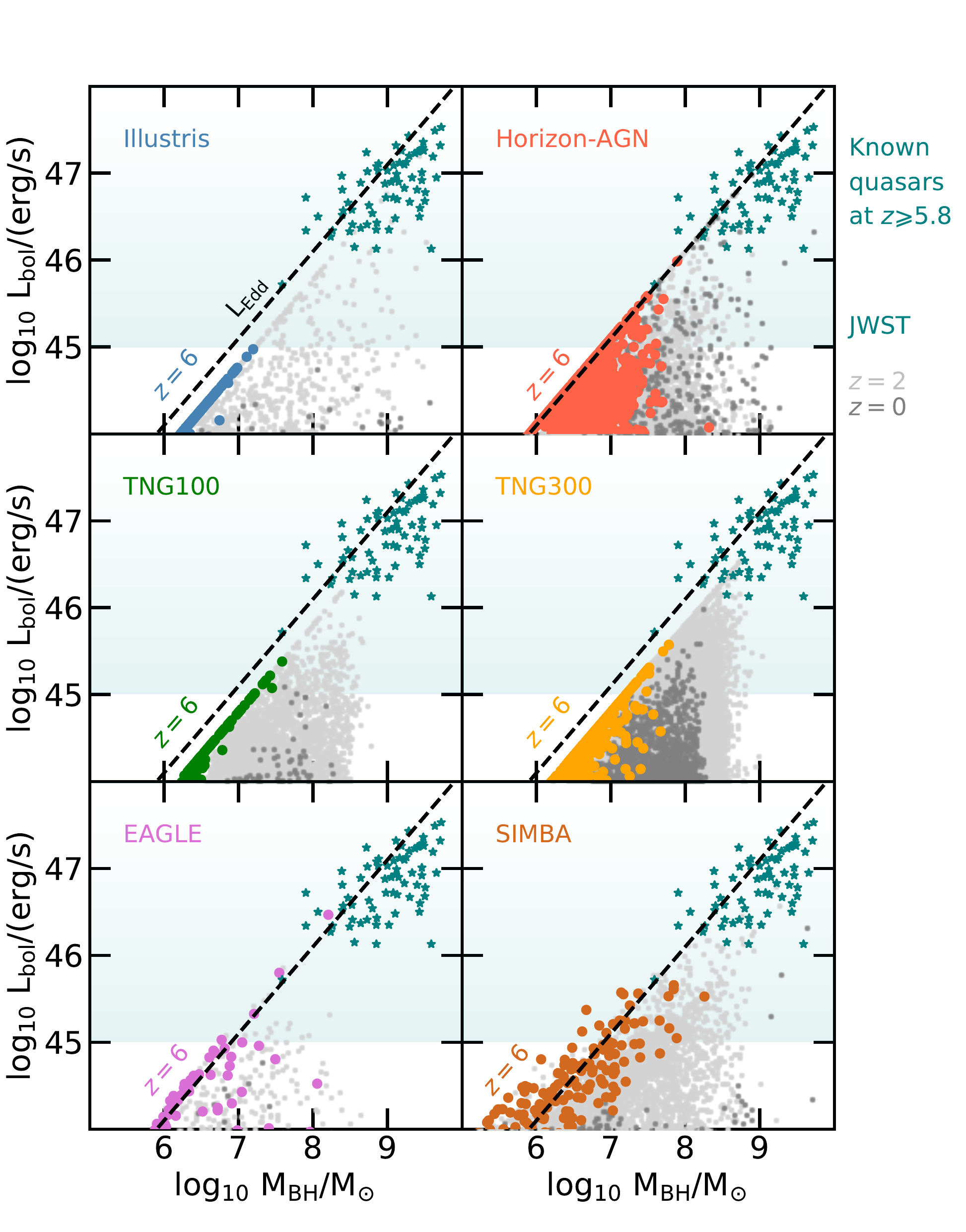}
    \caption{
    Bolometric luminosity of BHs produced by large-scale cosmological simulations as a function of their masses. Here we include BHs of $M_{\rm BH}\geqslant 10^{5}\, \rm M_{\odot}$ (if available) located in galaxies of $M_{\star}\geqslant 10^{9}\, \rm M_{\odot}$.
    Each panel represents a different simulation. In colors (e.g., blue, green), we show the population of simulated BHs produced at $z=6$. For reference, we show the BH population at $z=2$ with light grey symbols, and the population at $z=0$ in dark grey.  The BH accretion rates in all the simulations except SIMBA are capped at the Eddington limit, and we find that at $z=6$ a significant fraction of BHs accrete at this limit in all the simulations. 
 The BHs of Illustris, and TNG100, TNG300 are below the Eddington limit because we employ a radiative efficiency of $\epsilon_{\rm r}=0.1$ in our analysis instead of the $\epsilon_{\rm r}=0.2$ used in these simulations.
SIMBA produce super-Eddington BHs, as BH accretion rates are not capped to the Eddington limit (the limit is shown with a black line). The Horizon-AGN, TNG300, EAGLE and SIMBA simulations have a scatter of $L_{\rm bol}$ at fixed $M_{\rm BH}$, while Illustris and TNG100 produce a very tight $L_{\rm bol}-M_{\rm BH}$ correlation. 
    The population of $z\geqslant 5.8$ quasars with BH mass measurements, mostly with $L_{\rm bol}\geqslant 10^{46}\, \rm erg/s$, is shown with green star symbols \citep[][and Onoue et al., in prep]{2018ApJS..237....5M,2019ApJ...880...77O}. JWST should be able to characterize fainter quasars with $L_{\rm bol}\geqslant 10^{45}\, \rm erg/s$ (green shaded region).}
    \label{fig:LbolBH_plane}
\end{figure*}

\begin{figure}
    \centering
    \hspace*{-0.3cm}
    \includegraphics[scale=0.5]{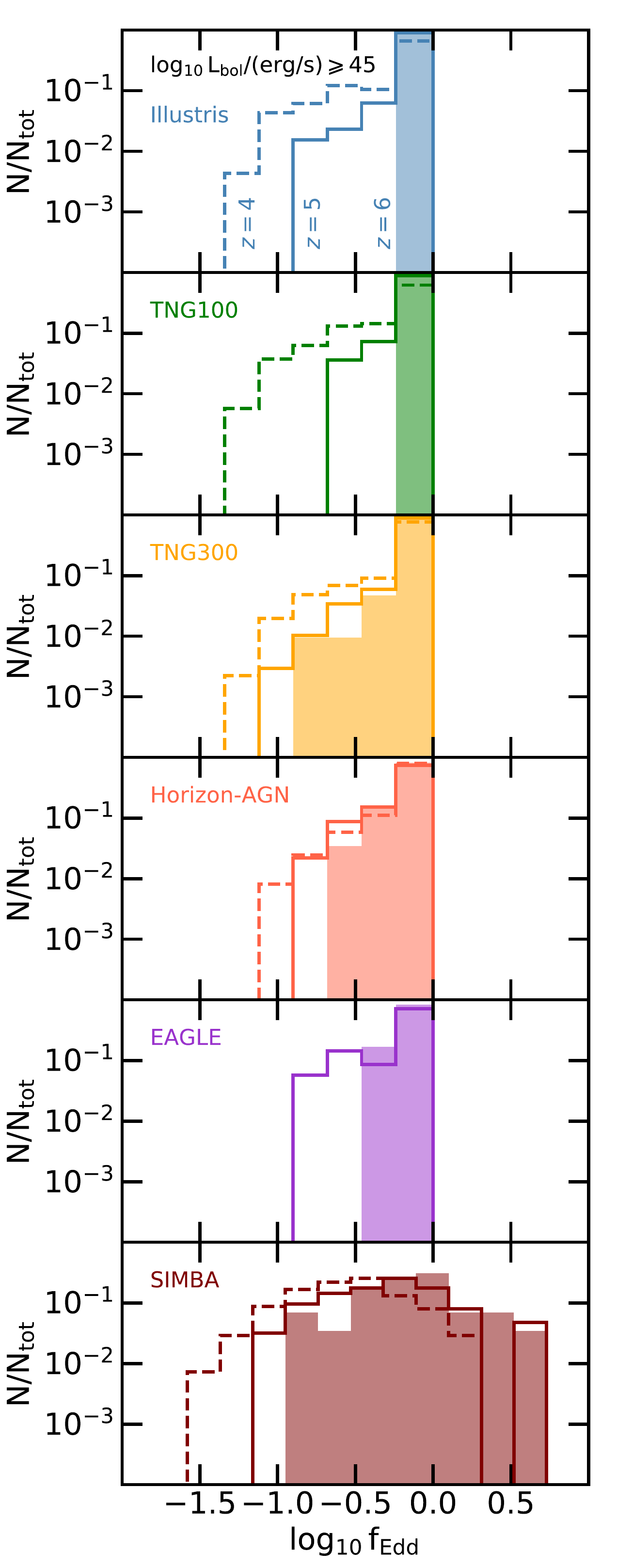}
    \caption{Eddington ratios of simulated quasars with $L_{\rm bol}\geqslant 10^{45}\, \rm erg/s$ at $z=6$ (shaded histograms), $z=5$ (solid lines), and $z=4$ (dashed lines).
    SIMBA is the only simulation allowing quasars to accrete above the Eddington limit.
    At $z=6$, most of the quasars in most of the simulations have Eddington ratios close to unity. We note that the TNG300, Horizon-AGN, and SIMBA produce Eddington ratio distributions extending to $\log_{10}\, f_{\rm Edd}\sim -0.75 \,{\rm or}\, -1.0$. With decreasing redshifts, all the distributions extend to smaller Eddington ratios.
    } 
    \label{fig:Edd_ratio}
\end{figure}

\subsection{Number density of AGN, faint and bright quasars}
To appreciate the number of AGN and quasars in cosmological simulations, we show in Fig.~\ref{fig:number_density} the number density of BHs with $\geqslant 10^{6}\,\rm M_{\odot}$ in three ranges of bolometric luminosity, in galaxies with stellar mass of $M_{\star}\geqslant 10^{9}\,\rm M_{\odot}$. AGN with $L_{\rm bol}=10^{44}-10^{45}\, \rm erg/s$ are shown in the left panel, and have number densities ranging in $n_{\rm AGN}=10^{-5}-10^{-3}\, \rm cMpc^{-3}$. In the middle panel we show the number density of {\it faint} quasars with $L_{\rm bol}=10^{45}-10^{46}\, \rm erg/s$. In this range, the simulations predict $n_{\rm faint\, quasars}=10^{-6}-10^{-5}\rm \, cMpc^{-3}$ at $z=6$.
Finally, we show brighter quasars with $L_{\rm bol}\geqslant 10^{46}\, \rm erg/s$ in the right panel. Unfortunately for this latter sub-population of BHs, these large-scale simulations do not allow us to have robust statistics at $z=6$ with number densities of $n_{\rm bright\, quasars}\leqslant 10^{-6}\, \rm cMpc^{-3}$. 
For example, in SIMBA there are only 7 such quasars at $z=5.5$, 1 quasar at $z=5$ in Illustris, 1 quasar at $z=6$ in EAGLE and in Horizon-AGN, 6 in TNG300 at $z=5$, and none in
TNG100. This is due to a combination of the underlying galaxy formation physics models and the limited volumes of these simulations, which span $100^{3}-300^{3}\, \rm cMpc^{3}$. Statistics increase in the redshift range $z=3-5$. 
Therefore the results that we present below for these bright quasars lack statistically robust sample sizes.
This also motivates the need for simulations such as Illustris, TNG, Horizon-AGN, EAGLE, and SIMBA but with larger volumes.

Three aspects are noticeable in Fig.~\ref{fig:number_density}. First, there is an overall decrease in number densities for brighter objects. 
Second, the number density of the objects increases from high to low redshifts, reaches a peak, and decreases.  
Observations suggest that the number density of luminous AGN increased from the early Universe to $z\sim2$ and then decreased to the current Universe \citep[e.g.,][]{Croom2004,Barger2003b,Richards2006,Matute2006,2009MNRAS.399.1755C,2010MNRAS.401.2531A,2013pss6.book..503M}. All the simulations presented here reproduce this behavior: the number density of AGN or quasars increases from $z=6$ to $z=3-2$ and later decreases to the present-day Universe.
However, the exact redshift at which the number density peaks varies from one simulation to another. For the AGN with $\log_{10}\,L_{\rm bol}/(\rm erg/s)=44-45$, the peak appears at $z\sim 1$ in SIMBA, at $z\sim 2$ in TNG100, TNG300, and EAGLE, and at earlier times $z\sim 3$ in Horizon-AGN and Illustris.

The third effect that one can notice is that the redshift at which the object number density peaks depends on the luminosity of the objects, as shown in Fig.~\ref{fig:number_density}.
In several simulations the {\it downsizing} effect is clear: bright objects peak at earlier time than fainter ones. In SIMBA, the number density of bright quasars (right panel) peak at $z\geqslant 2$ and the AGN (left panel) at $z\sim 1$. In TNG, bright quasar number density peaks at $z\sim 3$, and AGN at $z\sim 2$. In the other simulations, the downsizing effect is not obvious.
The downsizing effect has been found in other cosmological simulations \citep[e.g.,][]{2014MNRAS.442.2304H}, and semi-analytical models \citep[e.g.,][]{2012MNRAS.419.2797F,Hirschmann2012}

In addition to the different aspects discussed above, Fig.~\ref{fig:number_density} shows that there is no consensus on the number density of AGN or quasars at fixed redshift in the simulations. This echoes the differences identified in the AGN luminosity function, especially for  $z\leqslant 4$, for the Illustris, TNG, Horizon-AGN, EAGLE, and SIMBA simulations (Habouzit et al., submitted).

\subsection{The BH populations that power the faint and bright quasars}
We show in Fig.~\ref{fig:LbolBH_plane} the bolometric luminosity of the simulated BHs as a function of their masses for the Illustris, TNG100, TNG300, Horizon-AGN, EAGLE, and SIMBA simulations. We particularly highlight the simulated populations of BHs at $z\sim 6$ with colored dot symbols. All the simulations but SIMBA are capped at the Eddington limit, SIMBA allow accretion rates larger than the Eddington limit. A significant fraction of the simulated BHs with $M_{\rm BH}\geqslant 10^{6}\, \rm M_{\odot}$ have Eddington luminosities, and appear on a linear relation in the $L_{\rm bol}-M_{\rm BH}$ plane (shown as black dashed lines in Fig.~\ref{fig:LbolBH_plane}).
For reference, we show the same plane at $z=2$ and $z=0$: for most of the simulations the $z=6$ small $L_{\rm bol}$ scatter grows with time (at least down to $z=2$) towards less luminous objects \citep[][for a complete analysis of the simulated AGN populations]{2021MNRAS.tmp.2905H}.
In SIMBA, several BHs at $z=6$ have a bolometric luminosity larger than their Eddington luminosity, and are therefore super-Eddington BHs. In the torque model of SIMBA's accretion subgrid model, BHs are allowed to accrete at rates up to three times larger the the Eddington limit \citep{2019MNRAS.486.2827D,2019MNRAS.487.5764T}.
The luminosity of the AGN in all the other simulations is capped at the Eddington limit.
We note that the Illustris and TNG100 simulations have a very tight $L_{\rm bol}-M_{\rm BH}$ relation in Fig.~\ref{fig:LbolBH_plane}, while the TNG300, Horizon-AGN, EAGLE and SIMBA simulations produces a larger scatter of $L_{\rm bol}$ luminosities at fixed BH mass. Comparing the different simulations is difficult here, as they do not use the same volume, resolution, seeding and accretion subgrid models.
In the case of the TNG simulations, the larger volume and the lower resolution of TNG300 are responsible for the larger $L_{\rm bol}$ scatter at fixed $M_{\rm BH}$. While the larger volume of TNG300 ensures more diversity of BH environments, its lower resolution resolves less accurately BH surroundings. This leads to lower gas densities, and thus to lower accretion rates onto the TNG300 BHs. The impact of the SN feedback is also stronger in TNG300 than in TNG100, leading again to a larger scatter of bolometric luminosity at fixed BH mass.

We add to Fig.~\ref{fig:LbolBH_plane} with green star symbols the quasars that have been observed at $z\geqslant 5.8$ with BH mass measurements from \citet{2019ApJ...880...77O} and Onoue et al. (in prep). The observed quasars have bolometric luminosities of $L_{\rm bol}=10^{46}-10^{47.5}\, \rm erg/s$, with one quasar with $M_{\rm BH}\sim 10^{7.6}\, \rm M_{\odot}$ and a slightly lower luminosity of $L_{\rm bol}\sim 10^{45.7}\, \rm erg/s$. At fixed BH mass, these quasars have a $L_{\rm bol}$ scatter\footnote{Each observational data point in Fig.~\ref{fig:LbolBH_plane} also carry a 0.5 dex systematic uncertainty due to the MgII-based BH measurements \citep{2013BASI...41...61S}.} of about 0.5 to 1 dex, similar to the scatter found in some simulations such as SIMBA.
At such high redshift, observations are currently strongly biased towards the brightest objects. While limited by their volumes, some of the simulations produce objects that are similar to the faintest of the observed quasars. Horizon-AGN has two BHs with $M_{\rm BH}=10^{7.5}\, \rm M_{\odot}$ and $L_{\rm bol}=10^{45.6}\, \rm erg/s$. 
TNG300 has two of those quasars powered by slightly more massive BHs ($M_{\rm BH}=10^{7.7}, 10^{7.8}\, \rm M_{\odot}$). EAGLE has one similar BH, but also one very bright and massive BH of $M_{\rm BH}=10^{8.2}\, \rm M_{\odot}$ and $L_{\rm bol}=10^{46.5}\, \rm M_{\odot}$. EAGLE and Horizon-AGN are the only simulations studied here with a quasar entering the $L_{\rm bol}-M_{\rm BH}$ region covered by the observed bright quasars with $L_{\rm bol}\geqslant 10^{46}\,\rm erg/s$. This is interesting for EAGLE, because its AGN population at later times ($z=5-0$) is, on average, fainter than the populations of the other simulations, as indicated by the dark ($z=0$) and light ($z=2$) grey dot symbols. 
SIMBA is different and produces seven BHs of $L_{\rm bol}=10^{45.6}-10^{45.7}$ at $z=6$ with different masses in the range $M_{\rm BH}=10^{7.1}-10^{8.3}\, \rm M_{\odot}$.

To quantify the accretion rates of the quasars with $L_{\rm bol}\geqslant 10^{45}\, \rm erg/s$, we show their Eddington ratios in Fig.~\ref{fig:Edd_ratio} (using the $\epsilon_{\rm r}$ used in each simulation).  In most of the simulations at $z=6$, these quasars have Eddington ratios close to unity. As shown in Fig.~\ref{fig:LbolBH_plane} some diversity exists in TNG300, Horizon-AGN, EAGLE and SIMBA, with some quasars having lower, but still high, Eddington ratios.

In the coming years the JWST telescope will be able to characterize faint quasars with $L_{\rm bol}\geqslant 10^{45}\, \rm erg/s$, shown as a green region in Fig.~\ref{fig:LbolBH_plane}, for which no BH properties measurements exist yet.
The simulations present diverse populations of quasars with bolometric luminosities in the JWST range, as shown by the number density of BHs with $L_{\rm bol}=10^{45}-10^{46}\, \rm erg/s$ in Fig.~\ref{fig:number_density} (middle panel). In the simulations, these faint quasars are powered by BHs of $M_{\rm BH}\geqslant 10^{7}\, \rm M_{\odot}$ and accreting at rates close to or at the Eddington limit (or above for SIMBA) as shown in Fig.~\ref{fig:LbolBH_plane}. In some simulations such as SIMBA or Horizon-AGN the population of faint quasars is more diverse in term of BH mass and luminosity: the faint quasars are not all accreting at the Eddington limit, but can accrete above the Eddington limit (for SIMBA) or below (SIMBA, Horizon-AGN).

One crucial question is whether these fainter quasars studied by JWST could be used to constrain the co-evolution of the BHs and their host galaxies at high redshift. We test this in the following section by deriving the $M_{\rm BH}$ mass offsets relative to the $M_{\rm BH}-M_{\star}$ relation at $z=0$.

\begin{figure*}
    \centering
    \hspace*{-0.438cm}
    \includegraphics[scale=0.973]{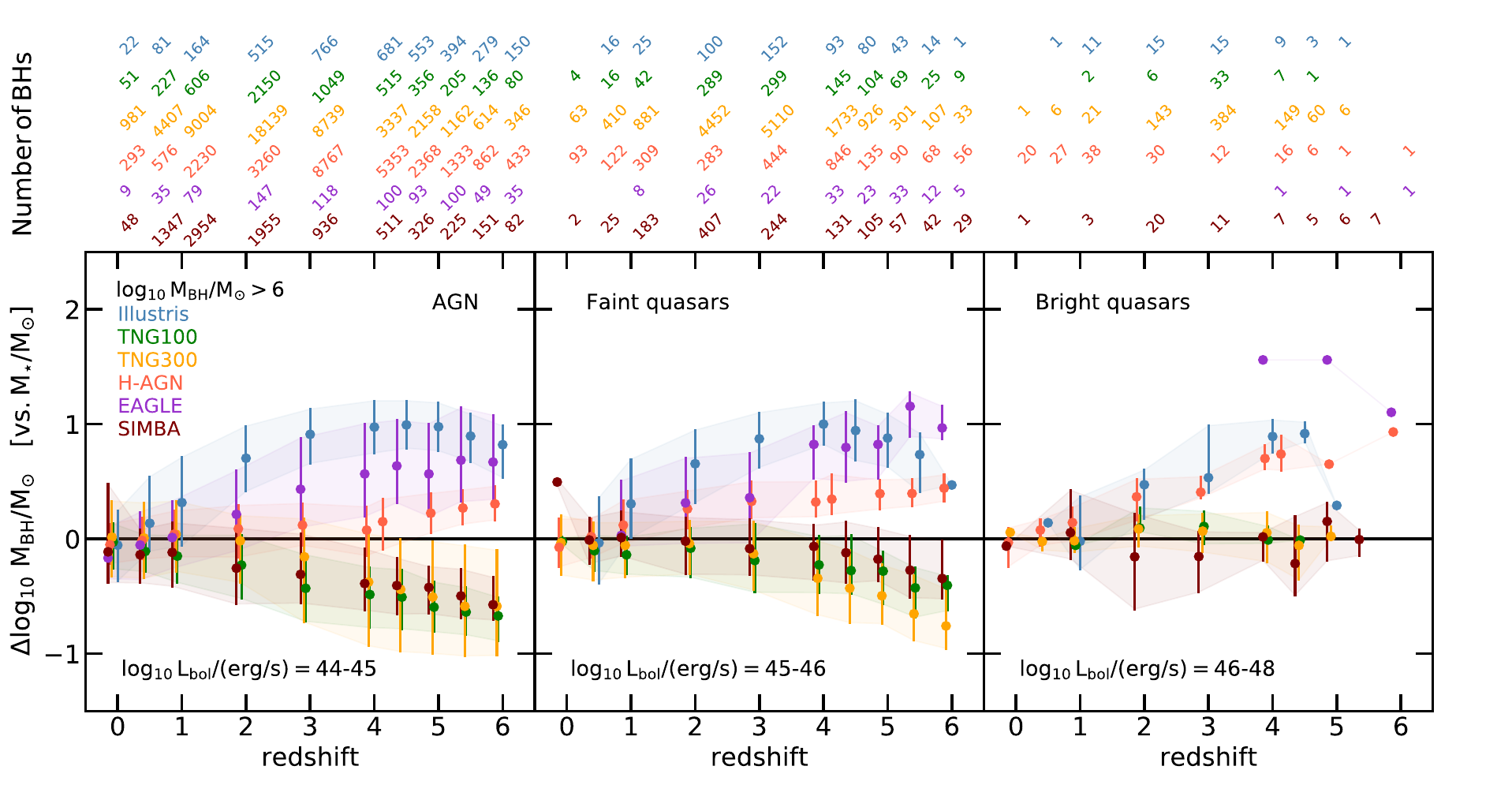}
    \caption{
    $M_{\rm BH}$ offset with respect to the mean $M_{\rm BH}-M_{\star}$ relation produced by the simulations at $z=0$, as a function of redshift. The 15th-85th percentiles of the offset distributions are represented with shaded areas, and the number of BHs at each redshift is indicated at the top of the figure. 
    {\it Left panel}: Offset for AGN with $L_{\rm bol}=10^{44}-10^{45}\, \rm erg/s$, which represents the bulk of the BH population at high redshift.
    There is no agreement between the simulations on the positive or negative offset. The difference between the simulations is mild at $z=0-1$, but can be as large as 2 dex at $z\geqslant 2$. {\it Middle panel}: For all the simulations, a similar offset is found for the {\it faint} quasars of $L_{\rm bol}=10^{45}-10^{46}\, \rm erg/s$ that will be characterizable by the JWST. 
    {\it Right panel}: The offsets produced by the simulations for quasars of $L_{\rm bol}\geqslant 10^{46}\, \rm erg/s$ do not match the offsets found for the bulk of the BH populations. 
    Our results show that characterizing fainter quasars (middle panel) than those in reach of current instrumentation (right panel), can constrain the assembly of the bulk of the BH population.
    Small samples of 5 or 10 faint quasars are sufficient to test the six simulations whose predicted BH offsets at high redshift are different by $\sim$1 dex. Given the simulation offset distributions at $z\geqslant 4$, we find that a sample of at least 5 faint quasars is sufficient to distinguish statistically (95$\%$ confidence) offsets larger than 0.2 dex from null offsets; more than 10 objects are needed for $\leqslant 0.2$ dex offsets (e.g., SIMBA or Horizon-AGN simulations at $z=4-5$).
        } 
    \label{fig:overmassive}
\end{figure*}

\begin{figure*}
    \centering
    \hspace*{-0.3cm}
    \includegraphics[scale=0.5]{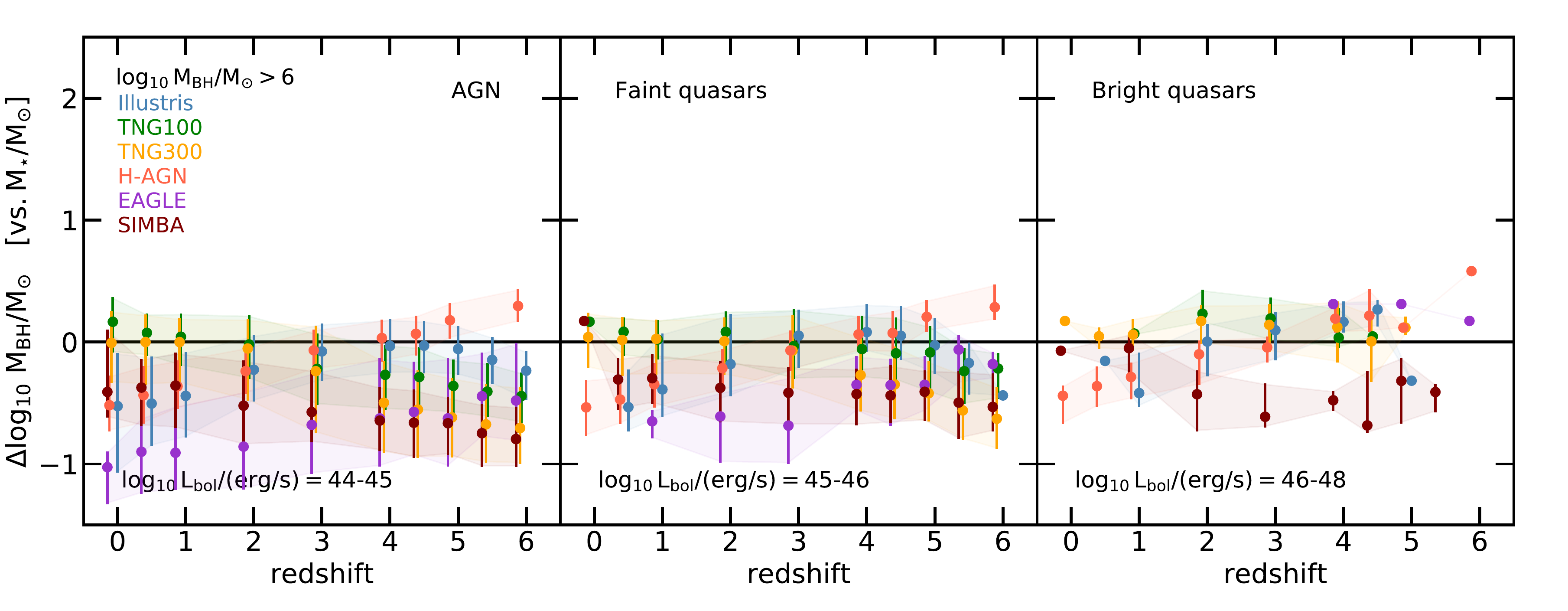}
    \caption{Same as Fig.~\ref{fig:overmassive} but here we use the local scaling relation of \citet{2013ARA&A..51..511K} to compute the $M_{\rm BH}$ offset. The differences between simulations are milder, but the main results are unchanged: faint quasars of $L_{\rm bol}=10^{45}-10^{46}\, \rm erg/s$ can constrain the $M_{\rm BH}$ offset of the BH population at high redshift, while the bright quasars in reach of current instrumentation may not be representative of the normal BH population.
   Observational samples of 10 objects (more than 10 objects) are needed to distinguish offsets larger than $0.2$ dex ($\leqslant 0.2$ dex) from null offsets, with $95\%$ confidence.
    }
    \label{fig:overmassive_Kormendy}
\end{figure*}

\begin{table*}
    \centering
    \caption{We define the BH mass offsets as $\Delta \log_{10} M_{\rm BH}=\log_{10} M_{\rm BH} (M_{\star}, z) - \log_{10} M_{\rm BH} (M_{\star}, z=0)$, and report in the table the median of the offsets, and the lower and upper limit of the offset distribution (standard deviation), predicted at $z=5$ and $z=6$ by the large-scale cosmological simulations. We only include the offsets for {\it faint} quasars with $M_{\rm BH}\geqslant 10^{6}\, \rm M_{\odot}$ and $L_{\rm bol}=10^{45}-10^{46}\, \rm erg/s$. For the first rows we use the simulation's $z=0$ scaling relation as a reference to compute the offsets (Fig.~\ref{fig:overmassive}), and for the second rows the \citet{2013ARA&A..51..511K} relation (Fig.~\ref{fig:overmassive_Kormendy}).}
    \begin{tabular}{l|c|c|c|c|c|c}
          & Illustris & TNG100 & TNG300 & Horizon-AGN & EAGLE & SIMBA \\
         \hline
         $z=6$ & 0.82 (0.53, 0.98) & -0.67 (-0.89, -0.51) & -0.59 (-1.01, -0.10) & 0.24 (0.088, 0.37) & 0.67 (0.34, 1.08) & -0.57 (-0.71, -0.34) \\
         $z=5$ & 0.98 (0.77, 1.18) & -0.59 (-0.81, -0.38) & -0.51 (-1.00, -0.024) & 0.13 (-0.04, 0.29) & 0.57 (0.20, 1.00) & -0.42 (-0.65, -0.24) \\
         \hline
         $z=6$ & -0.24 (-0.46, -0.09) & -0.44 (-0.65, -0.26) & -0.70 (-0.99, -0.38) & 0.31 (0.15, 0.43) & -0.48 (-0.81, -0.02) & -0.80 (-1.01, -0.54) \\
         $z=5$ & -0.06 (-0.26, 0.12) & -0.36 (-0.57, 0.15) & -0.62 (-0.94, -0.30) & 0.18 (0.03, 0.29) & -0.63 (-1.01, -0.14) & -0.67 (-0.92, -0.46) \\
    \end{tabular}
    \label{tab:offset_values}
\end{table*}

\begin{table*}
    \centering
    \caption{Difference offsets (in absolute values) between the median of the AGN mass offsets and the median of the faint and bright quasar offsets, for $z=4$ and $z=5$. Difference offsets are defined as: $\Delta {\rm off.\, faint} = \Delta \log_{10}\, M_{\rm BH}/\rm M_{\odot}|_{\rm faint} - \Delta \log_{10}\, M_{\rm BH}/\rm M_{\odot}|_{\rm AGN}$, and $\Delta {\rm off.\, bright} = \Delta \log_{10}\, M_{\rm BH}/\rm M_{\odot}|_{\rm bright} - \Delta \log_{10}\, M_{\rm BH}/\rm M_{\odot}|_{\rm AGN}$.
    These values can be inferred from Fig.~\ref{fig:overmassive} and Fig.~\ref{fig:overmassive_Kormendy}, for which the $M_{\rm BH}$ mass offsets are computed assuming the $z=0$ scaling relation produced by each of the simulations or assuming the \citet{2013ARA&A..51..511K}. Values of $\Delta {\rm off.\, faint}$ are always smaller than $\Delta {\rm off.\, bright}$, meaning that $M_{\rm BH}$ mass offsets of the faint quasar population at high redshift are always more representative of the AGN population offsets than the bright quasar population offsets. In some cases, the bright quasar offsets can be up to one order of magnitude higher/smaller than those of the AGN population, while faint quasar offsets are not higher/smaller than 0.3 dex the offsets than the AGN population.}
    \begin{tabular}{l|c|c|c|c|c|c|c|c}
     & \multicolumn{4}{c|}{reference for mass offset: simulation scaling rel. (Fig.~\ref{fig:overmassive})} & \multicolumn{4}{c}{reference for mass offset: empirical scaling rel. (Fig.~\ref{fig:overmassive_Kormendy})}\\
     & \multicolumn{2}{c|}{$z=4$} & \multicolumn{2}{c|}{$z=5$} & \multicolumn{2}{c|}{$z=4$} & \multicolumn{2}{c}{$z=5$} \\
     \hline
     Simulations & $\Delta {\rm off.\, faint}$ & $\Delta {\rm off.\, bright}$& $\Delta {\rm off.\, faint}$ & $\Delta {\rm off.\, bright}$&$\Delta {\rm off.\, faint}$ & $\Delta {\rm off.\, bright}$ &$\Delta {\rm off.\, faint}$  & $\Delta {\rm off.\, bright}$\\
     \hline
     Illustris       & 0.03 & 0.08 & 0.10 & 0.69 & 0.11 & 0.20 & 0.03 & 0.26 \\
     \hline
     TNG100          & 0.26 & 0.48 & 0.31 & - & 0.21 & 0.30 & 0.27 & -\\
     \hline
     TNG300          & 0.03 & 0.43 & 0.01 & 0.53 & 0.23 & 0.62 & 0.20 & 0.74\\
     \hline
     Horizon-AGN     & 0.24 & 0.62 & 0.17 & 0.42 & 0.03 & 0.16 & 0.03 & 0.06\\
     \hline
     EAGLE           & 0.26 & 0.99 & 0.26 & 0.99 & 0.28 & 0.94 & 0.28 & 0.94\\
     \hline
     SIMBA           & 0.32 & 0.40 & 0.25 & 0.57 & 0.22 & 0.17 & 0.26 & -\\

    \end{tabular}

    \label{tab:agn_faint_offsets}
\end{table*}

\subsection{Constraining whether BHs are over-massive with faint high-redshift quasars}

We now turn to analyze the average offset of BH mass at a given redshift (e.g., $z=6, 5, 4$) compared to the local $M_{\rm BH}-M_{\star}$ relation at $z=0$, i.e. defined as $\Delta \log_{10} M_{\rm BH}=\log_{10} M_{\rm BH} (M_{\star}, z) - \log_{10} M_{\rm BH} (M_{\star}, z=0)$. Such offsets diagnose whether BH growth is faster or slower than the assembly of the host galaxies at high redshift with respect to the local Universe. 
It is important to notice, however, that when we compare the results of different simulations based on the offset of the BH-galaxy mass relation from that at $z=0$, differences across models may arise because different simulations predict not only different growth of their BHs, but also different growth of the stellar mass of galaxies \citep{2021MNRAS.503.1940H}.

In Fig.~\ref{fig:overmassive} we show the median of the offset distributions and their 15th-85th percentiles for all the simulations. As in the previous section, the three panels represent distinct ranges of bolometric luminosities, with $L_{\rm bol}=10^{44}-10^{45}\, \rm erg/s$, $L_{\rm bol}=10^{45}-10^{46}\, \rm erg/s$ for faint quasars, and brighter quasars with $L_{\rm bol}\geqslant 10^{46}\, \rm erg/s$. 
As shown by the 15th--85th percentiles the offset distributions are broad. 
While the offsets are within 0.1 dex at low redshift ($z=0-1$), they are significantly larger at higher redshift and reach 0.5 to 1 dex for $z\geqslant 2$.
The median offset varies strongly from one simulation to another. 
The offset for BHs with $L_{\rm bol}=10^{44}-10^{45}\, \rm erg/s$ and $10^{45}-10^{46}\, \rm erg/s$ is positive (i.e., BHs are more massive) at high redshift for the Illustris, EAGLE\footnote{In both Fig.~\ref{fig:overmassive} and Fig.~\ref{fig:overmassive_Kormendy}, we note the absence of quasars with $\log_{10}\, L_{\rm bol}/({\rm erg/s})\geqslant 46$ in EAGLE for $z\leqslant 4$. This can also be noted from Fig.~\ref{fig:LbolBH_plane}. EAGLE has on average the faintest population of AGN among all the studied simulations, at any redshift, as shown by its luminosity function (Habouzit,submitted).}, and Horizon-AGN simulations, and negative (undermassive BHs) for the TNG100, TNG300, and SIMBA simulations. 
For clarity we do not show the Poisson errors in Fig.~\ref{fig:overmassive}, but we provide the numbers of objects $N_{\rm objects}$ for each redshift at the top of the figure. Poisson errors (offset/$\sqrt{N_{\rm objects}}$) indicate that the offsets at high redshift are statistically distinguishable from null offsets for all the simulations, even for small number of objects.
As discussed in Section 3.1, the positive offsets of Illustris and Horizon-AGN compared to the simulation's $z=0$ scaling relation are due to galaxies growing on average more, at lower redshift, than their BHs. The positive offsets of EAGLE at high redshift are due to efficient BH growth taking place in less massive galaxies at early times than at later times \citep{2018MNRAS.481.3118M}. For example, at $z=5$ a BH located in a galaxy of $10^{9.5}\, \rm M_{\odot}$ galaxy can grow efficiently, while another BH in a same-mass galaxy at $z=0$ would not. 
The negative offsets of the TNG simulations are mostly due a more efficient SN feedback at high redshift, which leads to lower average BH masses at fixed stellar mass than in the low-redshift Universe. In SIMBA, the additional torque model channel of BH accretion leads to more massive BHs with time, and thus negative offsets at high redshifts. We report the offsets obtained for the {\it faint} quasars in Table~\ref{tab:offset_values} for $z=5$ and $z=6$.

The evolution and values of the offsets for the two bolometric luminosity ranges $L_{\rm bol}=10^{44}-10^{45}\, \rm erg/s$ and $L_{\rm bol}=10^{45}-10^{46}\, \rm erg/s$ (i.e., AGN and {\it faint} quasars) are very similar. We emphasize here again that the offsets of the AGN population are almost identical to the offsets of all BHs (inactive and active) with $M_{\rm BH}\geqslant 10^{6}\, \rm M_{\odot}$, which justifies the use of the AGN sample as a proxy for the entire BH population (see Section~2.5).

The median offsets shown for the bright simulated quasars (right panel of Fig.~\ref{fig:overmassive}) are different from the AGN ones, especially at $z\geqslant 4$. The population of simulated BHs powering the bright quasars is consistent with or overmassive with respect to the simulation $z=0$ scaling relation, for all the simulations. This trend is consistent with recent ALMA observations, although using galaxy dynamical masses instead of stellar masses. We quantify this in Table~\ref{tab:agn_faint_offsets}: we compute the relative difference between the median offsets of the faint quasars and the AGN ($\Delta\, \rm off.\, faint$), and the median offsets of the bright quasars and the AGN ($\Delta\, \rm off.\, bright$), for $z=4$ and $z=5$.
The difference of faint quasar offsets to the AGN offsets is at most of $\sim 0.3$ dex at $z=4-5$ for all simulations, while they are up to 1 dex for the bright quasars.
Our results mean that observing {\it faint quasars} with $L_{\rm bol}=10^{45}-10^{46}\, \rm erg/s$ with JWST would provide us with offsets representative of a more {\it normal} population of BHs, here shown on the left panel of Fig.~\ref{fig:overmassive} with the $L_{\rm bol}=10^{44}-10^{45}\, \rm erg/s$ AGN. 
Our results also indicate that the extremely bright quasars observed in the Universe at $z\sim4-5$ are not representative of the {\it normal} BH population. Extrapolating our results to higher redshifts (for which current simulations do not produce enough objects), the observed bright quasars at $z\sim6$ would not be representative of the BH population either. 
Our work demonstrates that we have to be extremely careful when dealing with the observed high-redshift quasars: whether they are found to be overmassive or undermassive compared to local scaling relations, we can not extend these conclusions to the assembly of more {\it normal} BHs.

The BH mass offsets presented in Fig.~\ref{fig:overmassive} are relative to the mean $M_{\rm BH}-M_{\star}$ relation derived at $z=0$ in the simulations. These $z=0$ scaling relations are all different (see Fig.~\ref{fig:scaling}) and the time evolution of these relations significantly vary from simulation to simulation \citep{2021MNRAS.503.1940H}. These two aspects are responsible for the large discrepancies that we found in the BH mass offsets, and indicate that the assembly of BHs and their host galaxies with time is still uncertain today.
To test this further, we now derive the BH mass offsets relative to the \citet{2013ARA&A..51..511K} local scaling relation for all the simulations. We show the results in Fig.~\ref{fig:overmassive_Kormendy}.
The median offsets are completely different from Fig.~\ref{fig:overmassive}. The first feature to notice is the discrepancy at $z=0$: most of the simulations studied here have a mean $M_{\rm BH}-M_{\star}$ lying below the empirical \citep{2013ARA&A..51..511K} relation. The time evolution of the median offsets is the same as before, and the discrepancies among simulations is still large and of about 1 dex.
The most important result is that our previous conclusions are unchanged: while the offsets of {\it faint quasars} with $L_{\rm bol}=10^{45}-10^{46}\, \rm erg/s$ are the same as the offsets of $L_{\rm bol}=10^{44}-10^{45}\, \rm erg/s$, this is not the case for the bright quasars of the simulations ($L_{\rm bol}\geqslant 10^{46}\, \rm erg/s$).

Finally, we verify that our results on the offsets are robust for even a small sample of 10 faint quasars. To do so, we  randomly select several samples of 10 faint quasars, and find that their offsets were still representative of the offsets of the AGN sample. We also compute the number of quasars that need to be observed to statistically distinguish their offsets from null offsets at a $95\%$ confidence level. Given the simulation offset distributions, samples of at least 5 faint quasars are needed to distinguish offsets of $\geqslant 0.2$ dex at high redshift (as it is the case for most of the simulations), while more than 10 or 20 would be needed for $< 0.2$ dex offsets (e.g., SIMBA or Horizon-AGN at $z=4-5$).
This is promising as the JWST Cycle 1 General Observer program of \citet{2021jwst.prop.1967O} will, for example, target 12 faint quasars. Comparing the offsets from the simulations and the offsets that will be observed with JWST will require to first investigate the differences in the BH mass distribution of the simulated and observed BH samples. Any flux-limited observational sample is biased toward more massive BHs. This luminosity bias needs to be considered when trying to infer intrinsic offsets from observed ones \citep[see][]{2011A&A...535A..87S,2014MNRAS.438.3422S,2020ApJ...888...37D}.

\section{Discussion}

\subsection{Comparison with previous works and need for detailed analyses of the galaxy population at high redshift in simulations}
In this paper, we have shown the large diversity of BH mass offsets at fixed galaxy stellar mass, with respect to the $z=0$ BH-galaxy relation, for the Illustris, TNG100, TNG300, Horizon-AGN, EAGLE, and SIMBA simulations. There is no consensus on the offsets for these simulations, and this is also true for the MassiveBlackII simulation \citep{2015MNRAS.450.1349K}. MassiveBlackII has a similar volume as those studied here with $100 h^{-1} \rm Mpc$ side length. 
With this simulation \citet{2015MNRAS.454..913D} found a small average BH mass offset of $\Delta M_{\rm BH}=0.1-0.2$ (offset relative to the scaling $M_{\rm BH}-M_{\star}$ relation of the simulation at $z=0$, and computed for $M_{\rm BH}>10^{7}\, \rm M_{\odot}$) at $z=6$. This offset is smaller than any of the simulations studied in this paper. The average offset of \citet{2015MNRAS.454..913D} has a very mild evolution, and slightly diminishes with time from $z=6$ to $z=0$. This means that the MassiveBlackII BHs grow, on average, along the same scaling relation for their entire life. This is not the case in the Illustris, TNG, Horizon-AGN, EAGLE, and SIMBA simulations, for which we find an evolution of their mean $M_{\rm BH}-M_{\star}$ relation with time.

The time evolution of the mean $M_{\rm BH}-M_{\star}$ relation depends on both the growth of BHs with time and the growth of their host galaxies with time. As shown in \citet{2021MNRAS.503.1940H}, the evolution with time and shape of the relation can be strongly affected by galaxy physics (e.g., SN feedback) and galaxy evolution.
For example, in the Illustris and Horizon-AGN simulations the overall decrease of the scaling relation with time can be linked to galaxies growing faster than their BHs, and overproducing the galaxy stellar mass function at $z=0$.
Therefore, it appears important to test whether general properties of galaxies in cosmological simulations reproduce observations at high redshift, in our case beyond $z=4$. 
Among these general properties, the galaxy stellar mass function, the stellar to halo mass ratios, the UV luminosity function  \citep{2020MNRAS.492.5167V,2020MNRAS.495.4747S,2021arXiv210412788S}. Most simulations do not reach a consensus at $z=0$ for these properties, but investigating those at higher redshift is crucial.

\subsection{Agreement between the simulated populations of AGN and observations}
Most of the simulations studied in this paper produce a bolometric luminosity function that overestimates the current observational constraints from e.g. \citet{2006ApJS..163...50H} for $L_{\rm bol}\leqslant 10^{45.5}\, \rm erg/s$ and some of them do so even up to $L_{\rm bol}\leqslant 10^{46}\, \rm erg/s$ at $z=4$ (Habouzit et al., submitted). At this redshift, these AGN are distributed over two orders of BH mass ($M_{\rm BH}=10^{7}-10^{9}\, \rm M_{\odot}$) and galaxy mass \citep{2021MNRAS.503.1940H}. 
At higher redshift (e.g., $z\sim 6$), there is currently too few observations of faint quasars and too few simulated quasars in simulations to assess the agreement of the bolometric luminosity functions. 
The impact of a possible overestimation of the luminosity function in the simulations at $z=6$ on our results is difficult to evaluate, because the more numerous AGN would likely be distributed over several orders of BH and galaxy mass, as it is the case at $z=4$ in all the simulations.
We stress here again that despite the different AGN populations produced by the simulations, they all show the same offset signal for the AGN and the faint quasar populations, and a different signal for the bright quasars.

\subsection{Comparing simulations to high-redshift observations}
\subsubsection{Dynamical mass and stellar mass of galaxies}
A difficult aspect when comparing observations to simulations is to assess whether we actually compare the same quantities. Here, we look at BH mass offsets relative to the local $M_{\rm BH}-M_{\star}$ relation produced by all the simulations, with $M_{\star}$ the total mass of galaxies. In current observations of high-redshift quasars, 
stellar mass measurements are not possible and dynamical mass estimates from gas tracers are used instead (in most of the cases the [CII] 158\,$\mu$m line, \citealt[][]{2021arXiv210205679N}). 
Recently, \citet{2019MNRAS.488.4004L} investigated different tracers of galaxy mass in a single but high-resolution zoom-in simulation of a $z=7$ quasar. They showed that while the BH powering the quasar could appear overmassive using gas-based tracers employed to derive dynamical masses in unresolved observations, the quasar is not overmassive when using stellar-based tracers to derive the galaxy stellar mass. This work highlights the possibility that dynamical masses estimated through the virial theorem in observations could underestimate the actual dynamical mass of the quasar host galaxy systems, which would misleadingly imply overmassive quasars with respect to their hosts at high redshift. There has been progress in quasar host observations, resolving the gas kinematics and morphology \citep{2020A&A...637A..84P,2021arXiv210205679N}. In some observations the kinetic field can not be explained by assuming a simple thin rotating disk model. This can be due to AGN feedback perturbing the gas kinematics, or even due to the presence of a companion galaxy in the quasar host close environment \citep[as discussed in][]{2020A&A...637A..84P}, or because the disk is still not yet formed at $z\geqslant 6$ \citep{2021arXiv210101219M}.
In this respect, the JWST will provide the first opportunity to directly probe the host stellar content of high-redshift quasars.
The stellar masses of {\it faint} quasars derived from the JWST will serve us an ideal reference to be compared with cosmological simulations that we present in this paper.

\subsubsection{Gas and dust obscuration at high redshift}
In this work, we did not correct the intrinsic bolometric luminosity of the accreting BHs for gas and dust obscuration \citep[but see][]{2020MNRAS.495.2135N,2020MNRAS.499.3819M}.
The bolometric luminosities of the $z\geqslant 6$ observed quasars are computed using a UV to bolometric correction, and thus are also not corrected for dust obscuration.

While the observed quasars at $z\geqslant 6$ are non-obscured type I objects, large amount of dust have been measured in some high-redshift quasar hosts \citep[e.g.][]{2017ApJ...851L...8V,2019ApJ...881...63N,
2019ApJ...881L..23B,2020ApJ...904..130V}. While we can not completely exclude that the absence of dust obscuration correction in our work and/or in observations may lead to some mismatch, there is evidence for small obscuration in observations.
The UV obscuration of the known $z\geqslant 6$ quasars appears to be as small as lower redshift quasars \citep{2001AJ....121...54F,2001AJ....122..549V}, and indicates that the large amount of dust does not lead to significant obscuration along the line of sight.
This is another point where JWST can improve our current picture by constraining the dust extinction law probing the entire UV-MIR spectra of high-z quasars \citep[e.g.,][]{2020ApJ...905...51S,2021MNRAS.506.3946D}.

\subsection{Looking forward: Need for a diversity of larger volume cosmological simulations}
The BlueTides simulation is the largest simulation that was performed with only a slightly lower dark matter and gas resolution with respect to the simulations of volume $100^{3}-300^{3}\, \rm cMpc^{3}$ presented here. BlueTides has a volume $\geqslant 500^{3}\, \rm comoving \, Mpc^{3}$, but was only run down to $z=7$ \citep{2016MNRAS.455.2778F}.
The BlueTides simulation has a $M_{\rm BH}-M_{\star}$ relation at $z=7$ \citep{2020MNRAS.499.3819M} in agreement (but steeper) with the local scaling relation of \citet{2013ARA&A..51..511K}.
While BlueTides is useful to study high-redshift quasars, one can not derive the BH mass offsets relative to the simulation local ($z=0$) scaling relation as we did in this paper. Performing high-redshift larger volumes of Illustris, TNG, Horizon-AGN, EAGLE, and SIMBA, which have shown to all produce different population of BHs and AGN \citep[see also][Habouzit in prep]{2021MNRAS.503.1940H}, would allow us to investigate the quasar regime while already knowing how the subgrid physics shape the evolution of galaxies and BHs down to $z=0$ in less extreme regimes.
Because performing these simulations is computationally expensive, those could be run down to $z=6-5$ only.

The power of JWST to detect the stellar component of quasar host galaxies has been investigated with BlueTides at $z=7$ \citep{2021arXiv210101219M}. 
They find that the most massive simulated quasars ($M_{\rm BH}\sim 10^{8.4}-10^{8.9}\, \rm M_{\odot}$) are located in bulge-dominated galaxies which tend to be compact.
It is crucial now to assess what is the range of galaxy properties of the quasar hosts, as well as the properties of the BHs powering the quasars, produced by different models of galaxy formation and BH physics, i.e., for different large-scale cosmological simulations. 
The properties of the BH and AGN populations in current simulations \citep[e.g.,][Habouzit et al., sub.]{2019arXiv191000017L,2021MNRAS.503.1940H}, but also those of the galaxy population \citep[e.g.,][]{2017arXiv170302970P,2019ApJ...872..160H,2020arXiv201102501S}, are highly dependent on uncertain sub-grid model assumptions. Most of these models fail to capture the complex dynamics on small scales, i.e., 0.1-1000 pc \citep{2019MNRAS.483.3488B,2020arXiv200812303A}. Zoom-in cosmological simulations from larger volume cosmological simulations as mentioned above could help to achieve better resolution while capturing the quasars halo environment. This would allow to tackle key questions regarding the assembly of high-redshift quasars and their environments.

\section{Conclusions}
We analyzed the evolution of the BH population in six large-scale cosmological simulations: Illustris, TNG100, TNG300, Horizon-AGN, EAGLE, and SIMBA. We focused our analysis on the promising population of {\it faint quasars}, that we defined as active BHs with $L_{\rm bol}=10^{45}-10^{46}\, \rm erg/s$. They have the advantage to have better number statistics than brighter quasars in current cosmological simulations. Furthermore, both their host galaxy and BH properties will become characterizable by JWST.  
We analyzed how these quasars will yield new key constraints on the co-evolution of BHs and galaxies at high redshift.
We summarize our main findings below. 

\begin{itemize}
    \item In the large-scale cosmological simulations studied here there is no consensus on whether BHs at $z=6$ are overmassive or undermassive relative to either the simulation mean $M_{\rm BH}-M_{\star}$ relation at $z=0$ (Fig.~\ref{fig:scaling}, Fig.~\ref{fig:overmassive}), or the empirical scaling relation of \citet{2013ARA&A..51..511K} (Fig.~\ref{fig:overmassive_Kormendy}).

    \item In most of the simulations, BHs at $z=6$ are on average not as massive and bright as the quasars currently observed at the same redshift (Fig.~\ref{fig:LbolBH_plane}). 
    A significant fraction of the simulated massive $z=6$ BHs accrete mass at (or at rates close to) the Eddington limit (Fig.~\ref{fig:Edd_ratio}).
    The absence of BHs as massive as in the current observations is due to the limited volume probed by the simulations. Larger volume simulations would likely produce BHs overlapping with the observations, if accreting at the Eddington limit.

    \item Some simulations have a very tight $L_{\rm bol}-M_{\rm BH}$ relation for BHs of $M_{\rm BH}\geqslant 10^{6}\, \rm M_{\odot}$ and $\log_{10}\, L_{\rm bol}/( \rm erg/s)\geqslant 44$ at $z=6$, while some others produce a scatter of 0.5 dex or more in luminosity, at fixed BH mass. Such scatters are similar to high-redshift bright quasar observations. However, we note that this needs to be considered with caution as the simulated and observed populations are not for the same BH mass range.

    \item JWST will allow BH mass measurement of high-redshift {\it faint} quasars with $L_{\rm bol}\geqslant 10^{45}\, \rm erg/s$, going beyond of what is currently possible from the ground ($L_{\rm bol}\gtrsim 10^{46}\, \rm erg/s$; Fig.~\ref{fig:300quasars}).  
    This is a population of BHs that large-scale cosmological simulations produce in enough number for statistical analysis (Fig.~\ref{fig:LbolBH_plane}).
    
    \item At $z=6$, quasars with $L_{\rm bol}\geqslant 10^{45}\, \rm erg/s$ are in general among the most massive BHs present in the simulations at that time, with $M_{\rm BH}\geqslant 10^{7}-\rm a\, few\, 10^{8}\, \rm M_{\odot}$ (Fig.~\ref{fig:LbolBH_plane}). We find that in some simulations the quasars can also be powered by less massive BHs. The quasars are also not always the most massive BHs at fixed stellar mass. These differences among the simulations depend on the simulation subgrid physics, and particularly the accretion model. 
    \item  There is no consensus in the simulations on whether BHs are on average more, or less, massive at high redshift than at low redshift. Therefore, the BH mass offsets computed for the full BH population are crucial to understand the build-up of BHs at high redshift. 
    \item We find that the brightest BHs of $L_{\rm bol}\geqslant 10^{46}\, \rm erg/s$ at $z\leqslant 5$ do not trace the BH mass offsets of the full BH population (Fig.~\ref{fig:overmassive}). Extrapolating our results to $z=6$, the observed bright quasars could provide results not representative of the full BH population.
    However, simulated {\it faint} quasars with $L_{\rm bol}=10^{45}-10^{46}\, \rm erg/s$ (a range that JWST will be able to characterize) present the same BH mass offsets as the full BH population, at any redshift. The results are robust even with a small sample of 10 faint quasars. Moreover, we find that $\geqslant 0.2$ dex mean offsets would be distinguishable from null offsets for a sample of 10 observed faint quasars ($95\%$ confidence), while more quasars would be needed for smaller offsets. 
    High-redshift {\it faint quasars} will be key to constrain BH evolution at high redshift. 
    
    \item Large-scale cosmological simulations of $\geqslant 100 \, \rm cMpc$ side length are a great resource to study the evolution of the BH and galaxy populations, but are still lacking statistics for the most massive objects at $z\sim 6$. 
    Given the differences found in the simulations for the  faint $z\sim 6$ quasar population that JWST could characterize, we want to emphasize the need to run follow-up simulations (with the same subgrid physics) but with larger volumes. These simulations could be performed down to $z\sim 6$, to limit computational costs.  
\end{itemize}

In the coming years JWST will provide observations that can directly be compared with simulations and the results presented in this paper. For example, the approved Cycle 1 General Observer program by \citet{2021jwst.prop.1967O} is designed to provide a first detailed look at the central BHs and host galaxies of 12 of the lowest-luminosity quasars known at $z\sim 6$. 
This program will provide the H$\beta$-based BH masses and host stellar masses of their targets; therefore, it is expected that it will provide an observational test of the redshift evolution of the BH mass offset within the first billion years on the universe.
In the near future, the {\it Vera Rubin Observatory} and {\it Euclid} will provide us with a much large sample of $z\geqslant 6$ quasars that can be characterized in detail with JWST.
 
Our work also has important implications for the subgrid models that we employ in cosmological simulations. Since different simulations predict different BH mass offsets at high redshift, new observations could help constraining 
a  key regime in large-scale cosmological simulations, namely BH and galaxy formation and evolution at $z\geqslant 5$.

\section*{Acknowledgments}
We thank the referee for constructive comments on our paper.
MN acknowledges support from ERC Advanced grant 740246 (Cosmic\texttt{\char`_}Gas).
DAA was supported in part by NSF grants AST-2009687 and AST-2108944, and by the Flatiron Institute, which is supported by the Simons Foundation.

\section*{Data Availability Statement}
The data from the Illustris and the TNG100 simulations can be found on their respective websites: https://www.illustris-project.org, https://www.tng-project.org. The data from the EAGLE simulation can be obtained upon request to the EAGLE team at their website: http://icc.dur.ac.uk/Eagle/. 
The data from the SIMBA simulation can be found on the website: http://simba.roe.ac.uk/.
The Horizon-AGN simulation is not public, but some catalogs are available at: https://www.horizon-simulation.org/data.html.

\appendix
\section{Summary of the parameters and subgrid models employed in the simulations}
In Table~\ref{table:table_params}, we summarize the parameters employed in the simulations, and highlight their specific subgrid models. This table is a modified version of Table 1 of \citet{2021MNRAS.503.1940H}.
The last row indicates whether the $M_{\rm BH}-M_{\star}$ relation produced by the simulations increases or decreases with time, at fixed $M_{\star}$.

More detailed descriptions of the simulations and their BH modeling can be found in \citet{2014MNRAS.445..175G,2014MNRAS.444.1518V} for Illustris, \citet{2017arXiv170302970P,2018MNRAS.479.4056W} for TNG, \citet{2016MNRAS.463.3948D,2016MNRAS.460.2979V} for Horizon-AGN, \citet{2015MNRAS.446..521S,2016MNRAS.462..190R,2018MNRAS.481.3118M,2017MNRAS.468.3395M} for EAGLE, and \citet{2019MNRAS.486.2827D,2019MNRAS.487.5764T,2020arXiv201011225T} for SIMBA.

\begin{landscape}
\begin{table}
\caption{Parameters and models of BH and galaxy formation/evolution in the simulation Illustris, TNG100, TNG300, Horizon-AGN, EAGLE, and SIMBA. We include the quantities related to the volume and resolution of the simulations, the seeding prescriptions (i.e. minimum halo mass seeded, minimum cell density, velocity dispersion), and the BH mass of the seeds, the parameters of the BH accretion models (models and boost factors), the SN feedback models (models, efficiencies, and energy released per core collapse SN), and finally the parameters related to AGN feedback (number of modes, models, efficiencies, transition between modes).}
\begin{center}
\begin{tabular}{lcccccc}
\hline

\hline
& Illustris & TNG100 & TNG300 & Horizon-AGN & EAGLE & SIMBA\\
\hline

\hline
{\bf Cosmology} &&&&&\\
$\Omega_{\rm \Lambda}$ &0.7274&0.6911&0.6911&0.728&0.693&0.7\\
$\Omega_{\rm m}$ &0.2726&0.3089&0.3089&0.272&0.307&0.3\\
$\Omega_{\rm b}$ &0.0456&0.0486&0.0486&0.045&0.0483&0.048\\
$\sigma_{\rm 8}$ &0.809&0.8159&0.8159&0.81&0.8288&0.82\\
$n_{\rm s}$ &0.963&0.9667&0.9667&0.967&0.9611&0.97\\
$H_{\rm 0} \, \rm (km\, s^{-1}\, Mpc^{-1})$ &70.4&67.74&67.74&70.4&67.77&68\\
\hline
{\bf Resolution} &&&&&\\
Box side length ($\rm cMpc$)                                & 106.5 & 110.7 & 302.6 & 142.0 & 100.0& 147.1\\
Dark matter mass reso. ($\rm M_{\odot}$)     & $6.26\times 10^{6}$ & $7.5 \times 10^{6}$ & $5.9 \times 10^{7}$ & $8\times 10^{7}$ & $9.7\times 10^{6}$ & $9.6\times 10^{7}$\\
Baryonic mass reso. ($\rm M_{\odot}$)          & $1.26 \times 10^{6}$ & $1.4 \times 10^{6}$ & $1.1 \times 10^{7}$ & $2\times 10^{6}$  & $1.81\times 10^{6}$ & $1.82\times 10^{7}$\\
Spatial resolution ($\rm pkpc$)				   & 0.71 & 0.74 & 1.48 & 1.0 & 0.7 & 0.74\\
Gravitational softening ($\rm ckpc$)   & 1.4 	& 1.48 ($z\geqslant1$) & 2.96 ($z\geqslant1$)&	& 2.66 ($z\geqslant2.8$)	& 0.74\\
& & /0.74 pkpc & /1.48 pkpc	& & / max 0.7 pkpc &\\
Baryonic softening ($\rm ckpc$)       		& 1.4 ckpc ($z\geqslant 1$)	& 1.48 ($z\geqslant1$)	&2.96 ($z\geqslant1$)	&	& 2.66 ($z\geqslant2.8$) & 0.74	\\
& /0.7 pkpc 	&	/0.74 pkpc &	/1.48 pkpc &	&  / max 0.7 pkpc&	\\
\hline

{\bf Seeding} &&&&&\\
BH seed mass ($\rm M_{\odot}$)			  & $1.42\times 10^{5}$ & $1.18\times 10^{6}$ & $1.18\times  10^{6}$ & $10^{5}$ & $1.48\times 10^{5}$ & $1.43\times 10^{4}$\\
Seeding prescriptions                                           &  $M_{\rm h}/\rm M_{\odot}\geqslant$ & $M_{\rm h}/\rm M_{\odot}\geqslant$& $M_{\rm h}/\rm M_{\odot}\geqslant$ & $n\geqslant 0.1\, \rm H/cm^{3}$ & $M_{\rm h}/\rm M_{\odot}\geqslant$ & $M_{\star}/\rm M_{\odot}>$\\
&   									$7.1\times 10^{10}$ & $7.4\times 10^{10}$ & $7.4\times 10^{10}$ & $\sigma \geqslant 100\, \rm km/s$ & $1.48\times 10^{10}$ & $10^{9.5}$\\
\hline
Radiative efficiency $\epsilon_{\rm r}$                  & 0.2 & 0.2 & 0.2 & 0.1 & 0.1 & 0.1\\

\hline
{\bf Accretion} &&&&&\\
Model & Bondi  & Bondi + mag. field & Bondi + mag. field & Bondi  & Bondi + visc. & Bondi + torques\\
Boost factor & $\alpha=100$ & - & - & density-dependent  & -  & $\alpha=0.1$ \\
\hline

{\bf SN feedback}&&&&&\\
Model & kinetic & kinetic & kinetic & kinetic/thermal  & thermal & kinetic \\
\hline
{\bf AGN feedback} &&&&&\\
Single or 2 modes &2 modes &2 modes&2 modes&2 modes& single mode & 2 modes\\
High acc rate model & isotropic thermal  & isotropic thermal & isotropic thermal & isotropic thermal & isotropic thermal & kinetic \\
Feedback efficiency & $0.05\times0.2=0.01$ & $0.1\times0.2=0.02$ & $0.1\times0.2=0.02$ &$0.15\times0.1=0.015$& $0.1\times 0.15=0.015$ & $0.03\times0.1=0.003$\\

Low acc rate model & thermal hot bubble & pure kinetic winds & pure kinetic winds & kinetic bicanonical winds & -  & kinetic/ X-ray\\
Feedback efficiency & $0.35\times 0.2=0.07$ & $\leqslant0.2\times0.2= 0.04$ & $\leqslant0.2\times0.2= 0.04$ & $1\times0.1=0.1$ & - & $0.3\times0.1=0.03$ \\

Transition btw. modes  & $f_{\rm Edd}=0.05$ & $\min(0.002\left(\frac{M_{\rm BH}}{10^{8}\, \rm M_{\odot}}\right)^{2}, 0.1)$ & $\min(0.002\left(\frac{M_{\rm BH}}{10^{8}\, \rm M_{\odot}}\right)^{2}, 0.1)$ & $0.01$ & - &$0.2$ \\
\hline         

{\bf $M_{\rm BH}-M_{\star}$ evolution} & decrease & increase & increase & decrease & decrease & increase \\
{\bf with time} & at fixed $M_{\star}$ & at fixed $M_{\star}$&  at fixed $M_{\star}$ &  at fixed $M_{\star}$ & at fixed $M_{\star}$ &  at fixed $M_{\star}$\\

\hline
\end {tabular}
\end{center}
\label{table:table_params}
\end{table}
\end{landscape}

\bibliographystyle{mn2e}
\bibliography{biblio_complete_faintQSOs,biblio_complete,biblio_complete-2}

\label{lastpage}
\end{document}